\documentclass{elsart}
\usepackage{epsfig}
\usepackage{amssymb}
\usepackage{graphicx}
\usepackage{bm}
\begin{document}

\long\def\comment#1{ }

\newcommand{\BQ}{\begin{equation}}
\newcommand{\EQ}{\end{equation}}
\newcommand{\BQA}{\begin{eqnarray}}
\newcommand{\EQA}{\end{eqnarray}}
\newcommand{\be}{\begin{eqnarray}}
\newcommand{\ee}{\end{eqnarray}}
\newcommand{\NN}{\nonumber \\}
\newcommand{\del}{\partial}
\newcommand{\tr}{{\rm tr}}
\newcommand{\Tr}{{\rm Tr}}
\newcommand{\Path}{{\rm P}\,}
\newcommand{\ket}[1]{\left.\left\vert #1 \right. \right\rangle}
\newcommand{\bra}[1]{\left\langle\left. #1 \right\vert\right.}
\newcommand{\ketrm}[1]{\vert {\rm #1} \rangle}  
\newcommand{\brarm}[1]{\langle {\rm #1} \vert}  
\newcommand{\V}{\widetilde V}
\newcommand{\U}{\widetilde U}
\newcommand{\x}{\bm x}
\newcommand{\y}{\bm y}
\newcommand{\vv}{\bm v}
\newcommand{\z}{\bm z}
\newcommand{\w}{\bm w}
\newcommand{\q}{\bm q}
\newcommand{\kk}{\bm k}
\newcommand{\N}{{\mathcal N}_\tau}
\newcommand{\al}{\alpha}
\newcommand{\la}{\lambda}
\newcommand{\K}{{\mathcal K}}
\newcommand{\bb}{\bm b}
\newcommand{\rr}{\bm r}
\renewcommand{\theequation}{\thesection.\arabic{equation}}

\def\simge{\mathrel{%
   \rlap{\raise 0.511ex \hbox{$>$}}{\lower 0.511ex \hbox{$\sim$}}}}
\def\simle{\mathrel{
   \rlap{\raise 0.511ex \hbox{$<$}}{\lower 0.511ex \hbox{$\sim$}}}}
\def\bigs{\mathrel{
   \rlap{\raise 0.531ex \hbox{$>$}}{\lower 0.531ex \hbox{$<$}}}}
\def\buildchar#1#2#3{{\null\!
    \mathop#1\limits^{#2}_{#3}
    \!\null}}

\begin{flushright}
~\vspace{-1.25cm}\\
{\small\sf
 SPhT-T05/010\\ BNL-NT05/1
}
\end{flushright}
\vspace{1.cm}
\begin{frontmatter}

\title{Odderon in the Color Glass Condensate}

\author{Y. Hatta}
 \address{RIKEN BNL Research Center, Brookhaven National Laboratory,
Upton, NY 11973, USA}

\author{E. Iancu}\footnote{Membre du Centre National de la Recherche
Scientifique (CNRS), France.},
\author{K. Itakura}
\address{Service de Physique Th\'eorique, CEA Saclay, 91191
Gif-sur-Yvette, France}

\author{L. McLerran}
\address{Physics Department and RIKEN BNL Research Center, Brookhaven
National\\ Laboratory, Upton, NY 11973, USA}

\vspace{1.cm}


\begin{abstract}
We discuss the definition and the energy evolution of scattering
amplitudes with $C$-odd (``odderon") quantum numbers within the
effective theory for the Color Glass Condensate (CGC) endowed with the
functional, JIMWLK, evolution equation. We explicitly construct
gauge-invariant amplitudes describing multiple odderon exchanges in the
scattering between the CGC and two types of projectiles: a
color--singlet quark--antiquark pair (or `color dipole') and a system
of three quarks in a colorless state. We deduce the energy evolution of
these amplitudes from the general JIMWLK equation, which for this
purpose is recast in a more synthetic form, which is manifestly
infrared finite. For the dipole odderon, we confirm and extend the
non--linear evolution equations recently proposed by Kovchegov,
Szymanowski and Wallon, which couple the evolution of the odderon to
that of the pomeron, and predict the rapid suppression of the odderon
exchanges in the saturation regime at high energy. For the 3--quark
system, we focus on the linear regime at relatively low energy, where
our general equations are shown to reduce to the
Bartels--Kwiecinski--Praszalowicz  equation. Our gauge--invariant
amplitudes, and the associated evolution equations, stay explicitly
outside the M\"obius representation, which is the Hilbert space where
the BFKL Hamiltonian exhibits holomorphic separability.

\end{abstract}

\end{frontmatter}

\newpage


\section{Introduction}

Since the advent of the Balitsky--Fadin--Kuraev--Lipatov (BFKL)
equation \cite{bfkl,lipatov2} in the mid seventies, there has been
significant progress in our comprehension of high--energy QCD, and
several theoretical approaches have been proposed which aim at a
resummation of the energy--enhanced radiative corrections to
high--energy processes in perturbative QCD. The BFKL equation is a {\it
leading logarithmic approximation} (LLA), which allows one to resum to
all orders corrections of the form $(\alpha_s \ln s)^n$  to the
scattering between two colorless objects via the exchange of two gluons
in the $t$--channel. As a result of this resummation, the bare
two--gluon exchange is replaced by the {\it BFKL pomeron} (the sum of
an infinite series of ladder diagrams of ordinary perturbation theory),
or, equivalently, by two {\it reggeized} gluons which interact with
each other.
All the subsequent theoretical approaches proposed within perturbative
QCD encompass the BFKL equation, and can be viewed as extensions of the
latter towards increasing the complexity of the objects exchanged in
the $t$--channel, and also towards enlarging the limits of the LLA.

The simplest object beyond the BFKL pomeron within perturbative QCD is
the exchange of three interacting (reggeized) gluons in a symmetric
color state. This object, which is negative (or ``$C$--odd'') under
charge conjugation ($C=-1$),
represents the lowest order perturbative contribution to the {\it
odderon}, the $C$--odd exchange which dominates the difference between
the hadronic cross sections for direct and crossed channel processes at
very high energies \cite{nico}. The evolution of the three--gluon
odderon exchange with increasing energy in the LLA is described by the
BKP equation, established by  Bartels \cite{bartels} and Kwiecinski and
Praszalowicz \cite{pra}, which amounts to a pairwise iteration of the
BFKL kernel (see also \cite{Jaro}). This equation can be
immediately extended to describe the
exchange of an arbitrary number $n\ge 3$ of reggeized gluons with
pairwise BFKL interactions \cite{bartels0,bartels,lipatov,lipatov1,ko}.
The resulting formalism, also known as  the {\it generalized leading
logarithmic approximation} (GLLA), resums all radiative corrections
that involve the maximally possible number of energy logarithms $\ln s$
for a given number of exchanged gluons. At the moment, two exact
solutions of the BKP equation for odderon evolution are available
\cite{janik,BLV}, and the subject continues to be under intensive
debate \cite{ko2,de} (see also the recent review paper \cite{ewerz} and
the discussion below).

In the formalisms described so far,  the number of gluons in the
$t$-channel remains {\it fixed} in the course of the evolution. This is
probably a good approximation in some intermediate kinematical region,
but it fails to describe two interesting physical situations: First, it
does not incorporate correctly the {\it fluctuations in the number of
gluons}, as resulting from processes in which one (reggeized) gluon
splits into two, or, more generally, a $n$--gluon state evolves into a
$(n+m)$--one, with $m\ge 1$. Such processes are especially important in
the dilute regime at relatively large transverse momenta (for a given
energy), where gluon splitting is the main process through which
higher--point correlations get built 
\cite{it04}. Second, the approximation in which the number of
$t$--channel gluons is fixed cannot describe {\it recombination
processes} in which (reggeized) gluons merge with each other, thus
reducing the gluon density. Such processes are important in the
high--energy regime where the gluon density becomes large enough (due
to BFKL evolution and to the splitting processes alluded to above) to
enhance recombination processes, which are then expected to lead to
{\it gluon saturation} \cite{GLR,MQ86,MV94}. The inclusion of
saturation is also necessary, for consistency, in studies of the  {\it
unitarization} of the scattering amplitudes, except for some
exceptional kinematical configurations \cite{mueller}.

The simplest approach including {\it gluon splitting} in the framework
of BFKL evolution is the {\it color dipole picture} developed by
Mueller \cite{mueller,MP94}. This picture is valid at large $N_c$, and
describes pomeron multiplication via vertices at which one (BFKL)
pomeron splits into two. A more ambitious program, which is not
restricted to the large--$N_c$ approximation, is the {\it extended
generalized leading logarithmic approximation} (EGLLA), initiated by
Bartels \cite{bartels1}, in which the gluon number changing vertices
are explicitly computed in perturbative QCD (see Refs.
\cite{BW95,BV99,braun,BE99,ewerz0,bartels3,BE05} for further
developments along this line and  Ref. \cite{ewerz} for a review). By
using such vertices, evolution equations allowing for gluon splitting
have been written down in Refs. \cite{BV99,ewerz0,BE05}. Also, the
equivalence between the triple pomeron vertex in the dipole picture
\cite{MP94,RP97} and the one generated by EGLLA at large $N_c$
\cite{BW95,braun} has been verified in Refs. \cite{BV99,bartels3}.

 \comment{In principle, once these vertices are explicitly
known, they can be inserted in the appropriate evolution equations to
describe both splittings and mergings. In practice, it appears to be
difficult to describe mergings in this way, with the noticeable
exception of the Balitsky--Kovchegov (BK) equation \cite{balitsky,K},
that has been derived in this formalism too \cite{braun,bartels3}.}

So far, the only formalism allowing for the systematic inclusion of
{\it gluon merging} in the high--energy evolution is the {\it Color
Glass Condensate} (CGC) \cite{CGCreviews}, in which the reggeized gluons are
replaced by {\it classical color fields} whose correlations get built
in the course of the evolution. But the corresponding evolution is {\it
non--linear} : the new gluons radiated at one step in the evolution
(the analog of the `rungs' in the BFKL ladders) are allowed to scatter
off the classical color fields generated in the previous steps, and
this is the mechanism leading to gluon merging. Because of the
non--linear effects, the evolution couples $n$--point functions with
various values of $n$, and can be most compactly summarized as a {\it
functional Fokker--Planck equation} for the weight function describing
the correlations: the
Jalilian-Marian--Iancu--McLerran--Weigert--Leonidov--Kovner (JIMWLK)
equation \cite{jklw97,iancu,weigert}. Alternatively, and equivalently
\cite{path}, the evolution can be formulated as an hierarchy of
equations for scattering amplitudes
--- the Balitsky equations \cite{balitsky} ---, in which unitarity
is manifest. Note however that gluon {\it splittings} are not included
in the JIMWLK equation \cite{it04}; this is obvious from the fact that,
in the dilute, or {\it weak--field}, limit, this equation reduces to an
evolution in which the number of gluons in the $t$--channel stays {\it
constant} \cite{iancu,SAT}. An extension of the JIMWLK--Balitsky
evolution which includes pomeron splitting has been proposed only very
recently \cite{it04,MSW05}.

As it should be clear from this succinct presentation, the various
formalisms proposed so far in perturbative QCD at high energies are
quite different from each other, and the correspondences between them
are not always transparent. We know for instance that all these
approaches reproduce the  Balitsky--Kovchegov (BK) equation
\cite{balitsky,K}, which is the simplest non--linear generalization of
the BFKL equation, but only in the sense of a mean field approximation
that has been recently challenged \cite{it04,IMFLUCT,MS04,IMM04}. But
the relation between the {\it correlations} (i.e., the $n$--point
functions with $n
> 2$) generated by the different approaches is much less understood.
For instance, it has been shown only recently, by Kovchegov,
Szymanowski and Wallon \cite{kov}, that the perturbative odderon can be
accommodated within Mueller's dipole picture \cite{mueller}, and that
the corresponding solution coincides with the Bartels--Lipatov--Vacca
(BLV) solution \cite{BLV} to the BKP equation.

In particular, in the regime where saturation effects can be neglected,
one expects the CGC formalism and the more traditional approaches like
GLLA to be equivalent with each other, but this has never been verified
beyond the example of the 2--point function (i.e., of the BFKL
equation). With this paper, we would like to make one more step towards
elucidating this correspondence, by establishing the equivalence
between the two approaches at the level of {\it odderon} exchanges
(i.e., for a 3--point function). Specifically, we shall demonstrate
that, {\it in the weak--field limit}, the JIMWLK evolution of the
$C$--odd three--gluon exchanges reduces to the BKP equation, as
expected.

But recovering the BKP equation from the CGC formalism is not the main
purpose of the present analysis, but only a pretext for it. The CGC is
the theoretical framework par excellence for a study of high--energy
scattering and evolution in QCD near the unitarity limit, yet the
odderon problem has never been addressed in this formalism before.
Thus, a substantial fraction of the subsequent analysis will be devoted
to the proper formulation of the odderon exchanges in the framework of
the CGC, and to the derivation of the corresponding evolution equations
from the general, JIMWLK, equation. This study of the odderon should be
a good starting point towards understanding the multi--reggeon dynamics
within the CGC formalism.

Our study will also emphasize some essential differences between the
CGC formalism and the perturbative approach based on the BFKL
Hamiltonian: The latter is adapted to the description of a single
scattering via the exchange of a composite object --- pomeron, odderon,
or, in general, a system of $n$ reggeized gluons --- which evolves with
increasing energy. It relies on ``$k_T$--factorization'' (see, e.g.,
\cite{for}) to separate the dynamics in the transverse plane from that
in the longitudinal direction, and express a scattering amplitude as the
convolution of an universal {\it Green's function},
which describes the exchanged object, 
with the process--dependent {\it impact factors}, which connect this
object to the external particles. From the above, one sees that the
calculation is most naturally carried on in {\it momentum space}.

By contrast, in the CGC formalism
--- which is specially tailored to describe unitarity corrections ---,
single and multiple scatterings are treated on the same footing, namely
they are resummed in process--dependent, and gauge--invariant, {\it
scattering amplitudes}, which are computed in the {\it eikonal
approximation}, and thus are naturally constructed in {\it coordinate
space}. There is no $k_T$--factorization any longer, nor universal
Green's functions: the longitudinal and transverse dynamics are tied up
together in {\it Wilson lines}, which describe the eikonal scattering
of the elementary particles which compose the {\it projectile} (the
external object which scatters off the CGC, identified as the {\it
target}).

These differences explain some of the subtleties that we shall meet
when trying to compare results for the odderon in the two approaches.
On one hand, the odderon is described by the universal Green's function
of three reggeized gluons, which obeys BKP equation in momentum space.
On the other hand, the CGC scattering amplitudes depend upon the
specific process at hand (they include the impact factor of the
projectile) and obey non--linear evolution equations written in
coordinate space. (In general, these are not closed equations, but just
a part of Balitsky's hierarchy \cite{balitsky}.) Still, in the
weak--field, or single--scattering, approximation, in which the
evolution equations become linear, they must contain the same
non--trivial information as the BKP equation, whatever is the process
under consideration. 

The authors of Ref. \cite{kov} have met with a similar difficulty when
trying to compare the $C$--odd scattering amplitude of a dipole with
the standard BKP odderon. In that case, they have been able to do so by
using the respective solutions, which are explicitly known. Here, we
shall follow a more general strategy, which applies to arbitrary
processes, including those where the evolution equations are too
complicated to be solved exactly. Namely, by inspection of two specific
processes, we shall be able to identify the analog of the universal
odderon Green's function in the weak--field limit of the CGC formalism,
and show that, when properly defined, this quantity obeys indeed the
(coordinate version of the) BKP equation. As we shall momentarily
explain, this CGC approach to the BKP equation not only establishes a
correspondence between the two formalisms, but also reveals some new
insights about the BKP equation itself.

The two specific processes that we shall consider are the CGC
scattering with a {\it quark--antiquark color dipole} (a sub--process of
the virtual photon---CGC scattering) and that with a {\it colorless
3--quark system} (a simple model for a baryon). For both cases we start
by constructing the general, non--linear, amplitudes which describe
multiple odderon exchanges (these turn out to be the imaginary parts of
the respective $S$--matrix elements, themselves expressed in terms of
Wilson lines), and then expand these amplitudes in the limit where the
CGC field is weak. After this expansion, both amplitudes reduce to
(gauge--invariant) linear combinations involving a three--gluon Green's
function in a totally symmetric color state. Clearly, this Green's
function is a natural candidate for the BKP odderon in the CGC
formalism. This interpretation is, however, hindered by the fact that
the CGC Green's functions are gauge--{\it variant} objects, for which
the JIMWLK equation predicts infrared singularities (to be contrasted
with the BKP equation, which is infrared safe). Although physically
harmless --- as they cancel in the gauge--invariant amplitudes ---,
these singularities complicate the correspondence with the BKP
approach.

At this point comes one of the main new technical developments in this
paper: We show that the JIMWLK Hamiltonian \cite{iancu,weigert} can be
rewritten in a new form, which is {\it manifestly infrared finite} (the
original kernel in the transverse space is replaced by the dipole
kernel \cite{mueller}, which decays much faster at large distances).
When acting on gauge--invariant quantities, this new Hamiltonian is
equivalent with (but simpler to use than) the original one, in the
sense of generating the same evolution equations. But the new
Hamiltonian generates infrared--finite equations also for the
gauge--variant Green's functions, and thus allows us to introduce the
latter in a mathematically well--defined way. With this prescription,
the equation satisfied by the CGC odderon Green's function turns out to
be the same as the Fourier transform to coordinate space of the BKP
equation, as we shall check explicitly.

But this Fourier transform reserves some more surprises, as it could be
anticipated from the fact that our equation in coordinate space is {\it
not} exactly the same as the coordinate--space version of the BKP
equation that is usually written in the literature\footnote{At the
technical level, the difference originates in some ambiguities in the
form of delta--functions which appear when Fourier transforming the
momentum--space BKP equation to coordinate space, and which are
generally interpreted in the sense of the M\"obius representation
\cite{bartels3}; that is, these delta--functions are simply ignored.}
(see, e.g., \cite{ewerz}). Rather, the two equations coincide with each
other only if we require our CGC Green's function, which in general is
a totally symmetric function of three transverse coordinates, to vanish
whenever two coordinates become identical. This property is sometimes
referred to as ``the M\"obius representation'' (see, e.g.,
\cite{bartels3}), and is interesting in that, when restricted to
functions having this property, the BFKL Hamiltonian is conformally
invariant \cite{lipatov2} and exhibits holomorphic separability
\cite{lipatov}. This mathematical simplification has led
\cite{lipatov1,ko} to a powerful analogy between the BKP odderon
problem (and, more generally, the problem of multi--reggeon exchanges
in the limit of a large number of colors) and an integrable Heisenberg
spin chain. In particular, the first exact solution to the BKP equation
has been found, by Janik and Wosiek \cite{janik}, by exploiting this
analogy.

But although natural for the pomeron exchange (i.e., at the level of
the 2--point function), where it entails no loss of generality, the
restriction to the M\"obius representation is not so natural for the
higher $n$--point functions ($n\ge 3$), as recently emphasized in Ref.
\cite{bartels3}. For instance, the other known solution to the BKP
equation, due to Bartels, Lipatov, and Vacca \cite{BLV}, which
dominates at high energy and is perhaps of more relevance for the
phenomenology (as it couples to a virtual photon), lies outside the
M\"obius representation.

Similarly, the use of the M\"obius representation does not appear to be
natural in the CGC formalism either. In fact, our both examples of
gauge--invariant scattering amplitudes lie outside this representation:
For the dipole case, there is no coupling to this functional space, as
well known \cite{BLV,kov}, whereas for the 3--quark case, this property
is excluded by the initial conditions. We conclude that, at least for
the problems that we shall discuss, the BKP equation must be solved in
a Hilbert space more general than the M\"obius representation.

But our analysis below will not be confined to the weak--field limit and
its relation with the perturbative QCD approaches. As repeatedly
emphasized, the CGC is a formalism for multiple scattering, in which
non--linear amplitudes and the corresponding evolution equations are
straightforward to construct. As an illustration, we shall derive the
general evolution equations for the scattering amplitudes describing
$C$--odd and, respectively, $C$--even exchanges in the dipole--CGC
scattering (the non--linear generalizations of the odderon and,
respectively, pomeron exchanges). In fact, these equations will be
obtained by simply separating the real part and the imaginary part of
the first equation in the Balitsky hierarchy \cite{balitsky}.
Interestingly, the non--linear terms in these equations are found to
couple the odderon and pomeron evolutions. In the mean field
approximation in which the non--linear terms are assumed to factorize,
the equation for the $C$--odd amplitude reduces to a non--linear
equation originally proposed in Ref. \cite{kov}. Our analysis of this
equation will confirm the conclusion \cite{kov} that the odderon
exchanges are strongly suppressed by the unitarity corrections, and
will allow us to deduce the mathematical law for this suppression. For
the 3--quark system, we shall not write down the corresponding
non--linear equation (since this appears to be too complicated to be
illuminating). Rather, we shall rely on the relation between the
corresponding $C$--odd amplitude and the respective one for the dipole
to conclude that, in the weak--field regime, the dominant increase with
the energy should be controlled again by the BLV solution  \cite{BLV},
so like for the dipole case \cite{kov}.

The  plan  of  the  paper  is  as follows. In Sect. \ref{DIPOLE} we
give a general proof that the JIMWLK evolution of gauge--invariant
observables is free of infrared problems, and we deduce an alternative
form of the JIMWLK Hamiltonian which makes infrared finiteness
manifest. In Sect. \ref{WEAK_FIELD}, we consider the weak--field limit
of the JIMWLK evolution, and show that the use of the new Hamiltonian
allows one to introduce well--defined CGC Green's functions. In Sect.
\ref{OPERATORS} we construct the general amplitudes describing multiple
$C$--odd exchanges for a color dipole and a 3--quark system. Then, in
Sects. \ref{ODD_DIPOLE} and \ref{ODD_3Q}, we deduce the corresponding
evolution equations, after having introduced first the odderon Green's
function in the CGC. Finally, in Sect. \ref{Sect_BKP} we discuss the
connection to the BKP equation.

\section{The JIMWLK equation with the dipole kernel}
\setcounter{equation}{0} \label{DIPOLE}

In this section, we show that the JIMWLK equation can be considerably
simplified when its action is restricted to gauge--invariant
correlation functions, such as the scattering amplitudes. The resulting
equation is still a functional differential equation, but with a
different kernel in transverse space --- the {\it dipole kernel}
---, which has a more rapid fall--off at large distances, and
therefore makes it easier to check that the evolution is free of
infrared singularities.

\comment{Of course, for gauge--invariant quantities, the final
evolution equations come out the same whatever form of the functional
equation is used in their derivation. In such a case, the new,
`dipolar', form of the JIMWLK equation has merely the
(non--negligible) merits to simplify the algebraic manipulations and
render the intermediate steps manifestly infrared finite. But as we
shall argue in the next sections, the dipolar JIMWLK equation can also
be used for quantities which are not gauge invariant by themselves, so
like ordinary Green's functions, but which enter as building blocks in
the construction of the physical amplitudes\footnote{This is especially
useful in the weak--field regime at relatively low energy, where the
gauge--invariant amplitudes reduce to linear combinations of Green's
functions of the color fields.}. In such a case, the substitution of
the original JIMWLK kernel by the dipolar one is a convenient infrared
regularization allowing us to introduce well--defined Green's
functions, which is useful in intermediate calculations, and also for
comparison with more traditional approaches in perturbative QCD.}

\subsection{The JIMWLK equation}
\label{review}

The CGC formalism \cite{CGCreviews} is an effective theory for the
small--$x$ gluons in the light--cone wavefunction of an energetic hadron.
In this formalism, the gluons with small longitudinal momenta, or
small values of $x$, 
are described
as the classical color field radiated by `color sources' (gluons and
valence quarks) with higher values of $x$, which are `frozen'
by Lorentz time dilation in some random configuration. Accordingly, the
color fields at small--$x$ are themselves random, with a distribution
specified by the `weight function' $W_\tau[\alpha]$ (a functional probability
density). Here, $\tau \equiv \ln (1/x)$ is the rapidity, and
$\alpha\equiv \alpha^a(x^-,\x)$ is the light--cone component of the color
gauge field, and is the only non--trivial component in a suitable gauge
(the `covariant gauge'; see below). Note that this field depends upon the
light--cone longitudinal coordinate\footnote{We assume that the color glass moves
in the positive $z$ direction.} $x^-\equiv (t-z)/\sqrt{2}$, and upon the
transverse coordinates $\x=(x,y)$, but not  upon the (light-cone)
time $x^+\equiv (t+z)/\sqrt{2}$, in agreement with the `freezing' property
mentioned above.

All the interesting physical quantities
are expressed as operators built with $\alpha$, say
${\mathcal O}[\alpha]$, and the corresponding expectation values are
obtained after averaging over the random field $\alpha$:
\BQA
\left\langle {\mathcal O}\right\rangle_\tau \equiv \int
{\mathcal D}\alpha \, {\mathcal O}[\alpha]\, W_\tau[\alpha].
\label{averaging}
\EQA
Whereas, by itself, the weight function $W_\tau[\alpha]$ is a non--perturbative
object, its {\it evolution} with decreasing $x$ (or increasing energy) can be
computed in perturbation theory, at least in the high energy regime where
the intrinsic `saturation momentum' $Q_s(x)$ (which increases as a power
of $1/x$ \cite{GLR,DT02}) is hard. The corresponding evolution has
been computed in the non--linear generalization of the leading logarithmic
approximation \cite{jklw97,iancu}, which allows one to extend the BFKL
resummation \cite{bfkl} in the high density region at saturation. 
In this resummation, the radiative corrections enhanced by the logarithm
of the energy $\ln s \sim\ln 1/x$, and the non--linear effects involving the
classical field $\alpha$, are all treated on the same footing,
as effects of order one. The result of this calculation \cite{iancu}
is a second--order, functional, differential equation for $W_\tau[\alpha]$,
which is known as the JIMWLK equation \cite{jklw97,iancu,weigert} and reads:
\BQA \frac{\del}{\del\tau}W_\tau[\alpha]=HW_\tau[\alpha]\equiv
\frac12 \int_{xy}
 \frac{\delta}{\delta\alpha_\tau^a(\x)}\eta^{ab}(\x,\y)
 \frac{\delta}{\delta\alpha_\tau^b(\y)}W_\tau[\alpha],
\label{JIMWLK} \EQA where the subscript $xy$ on the integral sign
denotes the integration over the transverse coordinates $\x$ and $\y$.
The kernel $\eta^{ab}(\x,\y)$ is a functional of $\alpha$, upon which
it depends via the Wilson lines $\V(\x)$ and $\V^\dag(\x)$ built with
$\alpha\equiv \alpha^a T^a$ in the adjoint representation: \BQA
\eta^{ab}(\x,\y)=\frac{1}{\pi}\int\!\frac{d^2\z}{(2\pi)^2} \, {\mathcal
K}({\bm{x}, \bm{y}, \bm{z}})\,\left(1-\V^\dag_{\x}\V_{\z}\right)^{fa}
\left(1-\V^\dag_{\z}\V_{\y}\right)^{fb},
 \label{eta} \EQA
with the following transverse kernel:
\be\label{Kdef}
{\mathcal K}({\bm{x}, \bm{y}, \bm{z}})  \,\equiv \,
   \frac{(\bm{x}-\bm{z})\cdot(\bm{y}-\bm{z})}{
     (\bm{x}-\bm{z})^2 (\bm{z}-\bm{y})^2}\,,\ee
and, e.g.,
 \BQA \V^\dag_{\x}\equiv \V^\dag(\x)={\mbox P}\exp \left(
ig\int dx^-\alpha^a(x^-,\x)T^a \right), \label{up}
 \EQA
where P denotes path--ordering in $x^-$, and the integration over $x^-$
runs over the longitudinal extent of the hadron, which increases with
$\tau$ : When decreasing $x$, we include in the effective theory gluon
modes with smaller longitudinal momenta, which by the uncertainty
principle are localized at larger values of $x^-$. The functional
derivatives in Eq.~(\ref{JIMWLK}) act on the color field created in the
last step of the evolution (i.e., in the rapidity bin $(\tau,
\tau+d\tau)$), which is therefore located at the largest value of
$x^-$. Thus, the action of the  derivatives on Wilson lines like
(\ref{up}) reads as follows: \BQA \hspace*{-5mm}\frac{\delta
\V^\dag_{\x}}{\delta \alpha_{\tau}^a(\y)}=ig\delta^{(2)}(\x-\y)
T^a\V^\dag_{\x}\, , \quad \frac{\delta \V_{\x}}{\delta
\alpha_{\tau}^a(\y)}=-ig\delta^{(2)}(\x-\y)\V_{\x}T^a. \label{formulae}
\EQA

By taking a $\tau$-derivative in Eq.~(\ref{averaging}) and using the
JIMWLK equation (\ref{JIMWLK}), one can easily deduce the following
evolution equation for a generic observable:
\BQA
\frac{\del}{\del\tau}\left\langle {\mathcal O}
\right\rangle_\tau \,=\,\left\langle\frac12  \int_{xy}
 \frac{\delta}{\delta\alpha_\tau^a(\x)}\eta^{ab}(\x,\y)
 \frac{\delta}{\delta\alpha_\tau^b(\y)}{\mathcal O}
 \right\rangle_\tau.
\label{JIMWLK_evolution} \EQA For ${\mathcal O}$ to represent a
physical observable, this must be a {\it gauge--invariant} operator,
or, more precisely, the expression of such an operator when evaluated
in the covariant gauge in which $A^\mu_a(x)=\delta^{\mu
+}\alpha_a(x^-,\x)$. In such a case, one has seen on specific examples
that the evolution described by Eq.~(\ref{JIMWLK_evolution}) is
infrared safe \cite{iancu}, and in what follows we shall give a general
proof in that sense.

As a simple example, consider the scattering between the
CGC and an external `color dipole' : a quark--antiquark pair in a
colorless state. The corresponding $S$--matrix operator can be
computed in the eikonal approximation as:
 \BQA {\mathcal O}=\frac{1}{N_c}\tr(V^\dag_{\x} V_{\y}
)\equiv S(\x,\y;\alpha), \label{dipole} \EQA where $V_{\x}^\dag$ is a
Wilson line in the fundamental representation (as obtained by replacing
$T^a \to t^a$ in Eq.~(\ref{up})), and represents the phase factor
picked up by the quark while crossing the background field of the
target. Similarly, $V_{\y}$ is the corresponding phase factor for the
antiquark. Plugging this operator into Eq.~(\ref{JIMWLK_evolution}),
one  obtains \cite{iancu} the following equation (with
$\bar{\alpha}_s={\alpha_sN_c}/{\pi}$):
 \BQA \hspace*{-.6cm}\frac{\del}{\del \tau}\Big\langle \tr (V_{\x}^\dag
V_{\y})\Big\rangle_\tau =\frac{\bar{\alpha}_s}{2\pi} \!\int d^2\z
{\mathcal M}({\bm{x}},{\bm{y}},{\bm z}) \left\langle\frac{1}{N_c}\tr
(V^\dag_{\x} V_{\z})\tr (V^\dag_{\z} V_{\y}) - \tr (V^\dag_{\x} V_{\y})
\right\rangle_\tau, \label{Balitsky_fdt}\NN \EQA after a rather lengthy
calculation in which many terms which appear at intermediate steps
cancel with each other. This is not a closed equation, but only the
first equation in an infinite hierarchy originally derived by Balitsky
\cite{balitsky}. The kernel appearing in this integral has been
generated as: \be \label{Mdef} {\mathcal M}({\bm{x}},{\bm{y}},{\bm
z})\,\equiv\, \frac{(\bm{x}-\bm{y})^2}{(\bm{x}-\bm{z})^2
(\bm{z}-\bm{y})^2}\,= {\mathcal K}_{\bm{x} \bm{x} \bm{z}} + {\mathcal
K}_{\bm{y} \bm{y} \bm{z}} - 2{\mathcal K}_{\bm{x} \bm{y} \bm{z}} \,,\ee
and is recognized as the {\it dipole kernel} \cite{mueller}. Note that
the poles in this kernel at $\z=\x$ and $\z=\y$ are actually harmless
because the operator within the brackets vanishes at these points. In
fact, it is easy to check on Eq.~(\ref{eta}) that such `ultraviolet'
(i.e., short--distance) poles cancel already in the general evolution
equation (\ref{JIMWLK_evolution}), irrespective of the nature of the
operator ${\mathcal O}$.

The crucial new feature of the dipole kernel (\ref{Mdef}) as compared
to the original kernel (\ref{Kdef}) in the JIMWLK equation is that
this new kernel falls off much faster at large distances:
${\mathcal M}({\bm{x}},{\bm{y}},{\bm z})\sim 1/z^4$ when $z\gg x,\,y$,
which is enough to ensure the convergence of the integral in the r.h.s.
of Eq.~(\ref{Balitsky_fdt}). That is, this equation and, similarly,
all the higher equations in the Balitsky hierarchy
\cite{balitsky}, are {\it infrared safe}. As we shall explain
in the next subsection, this is related to the fact that the corresponding
operators are {\it gauge invariant}.

\subsection{The dipole JIMWLK equation and the finiteness conditions}

A brief inspection of the original JIMWLK equation, cf.
Eqs.~(\ref{JIMWLK})--(\ref{Kdef}), reveals that, for a {\it generic}
operator ${\mathcal O}$, there is {\it a priori} no guarantee that
the corresponding evolution equation (\ref{JIMWLK_evolution}) should be
infrared safe. Indeed, at large distances $z\gg x,\,y$,
the transverse kernel (\ref{Kdef}) decays only like
${\mathcal K}_{\bm{x} \bm{y} \bm{z}}\sim 1/z^2$, so
Eq.~(\ref{JIMWLK_evolution}) may  develop logarithmic divergences
($\sim \int d^2\z/\z^2$ for $\z\to \infty$). In the calculation
leading to (\ref{Balitsky_fdt}), such divergences have actually
canceled among various terms. In what follows, we shall argue that such
a cancellation will always hold for the {\it gauge--invariant} operators. In the
present context, where a gauge--fixing condition has been already chosen, what
we mean by that is the invariance under the residual gauge transformations.
An interesting consequence of this discussion will be that, for such
operators, the corresponding evolution equation (\ref{JIMWLK_evolution})
can equivalently be rewritten in a simpler, and manifestly infrared finite
form, which involves the dipole kernel (\ref{Mdef}):
\BQA
\frac{\del}{\del \tau}\langle {\mathcal O}\rangle_\tau
&=&\langle H_{{\rm dp}}{\mathcal O}\rangle \equiv
-\frac{1}{16\pi^3}\int_{xyz}\frac{(\x-\y)^2}{(\x-\z)^2(\z-\y)^2}
\label{JIMWLK_evolution_simple} \\
&&\qquad \quad \times\left\langle \left(1+\V^\dag_{\x}
\V_{\y}-\V^\dag_{\x} \V_{\z} - \V^\dag_{\z} \V_{\y} \right)^{ab}
\frac{\delta}{\delta\alpha^a_\tau(\x)}
\frac{\delta}{\delta\alpha_\tau^b(\y)}{\mathcal O}
\right\rangle_\tau.\nonumber \EQA

Our argument will be constructed as follows: First, we shall use Eqs.~(\ref{eta})
and (\ref{JIMWLK_evolution}) to deduce the ``finiteness conditions" that some
operator ${\mathcal O}$ must satisfy in order for the corresponding evolution
equation to be infrared safe. Then, we shall use these conditions to rewrite
Eq.~(\ref{JIMWLK_evolution}) in the form of Eq.~(\ref{JIMWLK_evolution_simple}).
Finally, we shall demonstrate that the ``finiteness conditions" are equivalent
to invariance under the residual gauge transformations in the covariant gauge.

The conditions for infrared finiteness are easily written down once one
realizes that the dangerous terms are those terms in Eq.~(\ref{eta})
which do not involve either $\V_{\z}$ or $\V^\dag_{\z}$. Indeed, after
rewriting $(1-\V^\dag_{\x}\V_{\z})(1-\V^\dag_{\z}\V_{\y}) =
1+\V^\dag_{\x} \V_{\y}-\V^\dag_{\x} \V_{\z} - \V^\dag_{\z} \V_{\y}$, it
becomes clear that the contribution of the first two terms,
$1+\V^\dag_{\x} \V_{\y}$, to Eq.~(\ref{JIMWLK_evolution}) is
potentially divergent: \BQA\label{diverg}
&&\int_{xyz}\frac{(\x-\z)\cdot(\y-\z)}{(\x-\z)^2(\z-\y)^2}
\frac{\delta}{\delta \alpha^a_\tau(\x)}(1+\V^\dag_{\x}\V_{\y})^{ab}
\frac{\delta}{\delta \alpha^b_\tau(\y)}{\mathcal O}\NN
&&\longrightarrow\quad \int\frac{d^2\z}{\z^2} \int_{xy}
\frac{\delta}{\delta \alpha^a_\tau(\x)}(1+\V^\dag_{\x}\V_{\y})^{ab}
\frac{\delta}{\delta \alpha^b_\tau(\y)}{\mathcal O}, \quad ({\rm for\
}\z\to \infty),\nonumber \EQA whereas the contributions of the other
terms should be finite, because correlations involving $\V_{\z}$ or
$\V^\dag_{\z}$ are expected to vanish when, e.g., $|\z-\x|\to\infty$.
Clearly, for the divergence in Eq.~(\ref{diverg}) to go away, the
coefficient of the divergent $\z$--integral there must vanish, which in
turn implies the two following {\it finiteness conditions}: \BQA &&\int
d^2\x \, \frac{\delta}{\delta \alpha^a_\tau(\x)}{\mathcal O}=0,
\label{finite_cond1}\\
&&\int d^2\x \, (\V_{\x})^{ab}
\frac{\delta}{\delta \alpha^b_\tau(\x)}{\mathcal O}=0.
\label{finite_cond2}
\EQA

It is easy to check that the following 2$n$-point operators constructed
from the Wilson lines in the fundamental representation \BQA {\mathcal
O}_n\equiv \tr \left( M_1 M_2 \cdots M_n \right),\qquad M_i\equiv
V^\dag_{x_i} V_{y_i}, \label{gauge_inv_op} \EQA (and arbitrary linear
combinations and products of them ${\mathcal O}_{n_1}{\mathcal
O}_{n_2}\cdots$) satisfy the finiteness conditions
Eqs.~(\ref{finite_cond1}) and (\ref{finite_cond2}). For example,
consider the simplest case $n=1$, i.e., the dipole operator
(\ref{dipole}). The first condition (\ref{finite_cond1}) is trivially
satisfied because
 \BQA \frac{\delta}{\delta \alpha_\tau^a(\vv)}\tr (V^\dag_{\x} V_{\y})
=ig\left\{\delta^{(2)}(\vv-\x) - \delta^{(2)}(\vv-\y)\right\}\,
 \tr(V^\dag_{\x} V_{\y} t^a)\nonumber \EQA
which indeed vanishes after integration over $\vv$, while the second
condition (\ref{finite_cond2}) can be verified with the help of the
formulae $(\V)^{ab}_v=2\tr(t^a V_v t^b V_v^\dag)$ and $\tr(t^a
A)\tr(t^a B)=\frac12 \tr(AB)-\frac{1}{2N_c}\tr(A) \tr(B)$.

Let us now establish the dipole form of the JIMWLK equation,
Eq.~(\ref{JIMWLK_evolution_simple}). We first decompose the original
kernel (\ref{Kdef}) so that the dipole kernel is separated out (cf.
Eq.~(\ref{Mdef})): \BQA 
\frac{(\x-\z)\cdot(\y-\z)}{(\x-\z)^2(\y-\z)^2} =\frac12 \left[
-\frac{(\x-\y)^2}{(\x-\z)^2(\y-\z)^2}+\frac{1}{(\x-\z)^2} +
\frac{1}{(\y-\z)^2} \right]. \nonumber \EQA Note that this is in fact
the separation of infrared divergences: The last two terms will
generate logarithmic divergences at large $\z$ when the JIMWLK
Hamiltonian is applied to a generic operator. However, it is easy to
show that if the operator satisfies the finiteness conditions, the
contributions coming from the last two terms ${1}/{(\x-\z)^2}$ and
${1}/{(\y-\z)^2}$ vanish. Thus, for such operators, we can replace
${\mathcal K}({\bm{x}},{\bm{y}},{\bm z}) \longrightarrow -(1/2)
{\mathcal M}({\bm{x}},{\bm{y}},{\bm z})$ in Eq.~(\ref{eta}), and at the
same time move the functional derivative
${\delta}/{\delta\alpha^a_\tau(\x)}$ to the right of the Wilson lines.
Indeed, the relevant commutator, namely, \BQA \hspace*{-6mm}\left[
\frac{\delta}{\delta\alpha^a_\tau(\x)}\,, \left(1+\V^\dag_{\x}
\V_{\y}-\V^\dag_{\x} \V_{\z} - \V^\dag_{\z} \V_{\y}\right)^{ab} \right]
=-ig \delta^{(2)}(\x-\y){\rm Tr} (\V^\dag_{\z} \V_{\y} T^b)
\label{commutator} \EQA vanishes when multiplied by the factor
$(\x-\y)^2$ in the dipole kernel. This establishes
Eq.~(\ref{JIMWLK_evolution_simple}).

Besides being conceptually more appealing (as infrared finiteness
is now manifest), the rewriting of the JIMWLK equation
as in Eq.~(\ref{JIMWLK_evolution_simple})
considerably simplifies the manipulations leading
from Eq.~(\ref{JIMWLK_evolution_simple})
to final evolution equations like Eq.~(\ref{Balitsky_fdt}).
The dipole kernel, that we expect in the final equations, is
already present there, and many of the terms generated at intermediate
steps when the calculations are based
on the original equations, Eqs.~(\ref{JIMWLK_evolution})--(\ref{eta}),
are simply absent when the calculations start with
Eq.~(\ref{JIMWLK_evolution_simple}).
For the more complicated operators that we shall encounter in the next
sections,
the usefulness of the dipole JIMWLK equation is indeed appreciable.

\subsection{Physical meaning of the finiteness conditions}

The operator (\ref{gauge_inv_op}) which has been seen
to obey an infrared--finite evolution equation
is the covariant--gauge expression of a gauge invariant operator.
Indeed,  one can rewrite this operator as the trace of
a closed Wilson loop, for which gauge symmetry is manifest. To that aim,
note that the end points at longitudinal infinity ($x^-=\pm\infty$)
of two adjacent Wilson lines (say, $V^\dag_{x_i}$
and $V_{y_i}$) can be connected by a Wilson line in the transverse
direction, which is simply unity in the present gauge, in which
$A^i_a=0$. Therefore, one can connect the end points of all the Wilson
lines which enter the trace in Eq.~(\ref{gauge_inv_op}) can be connected
in such a way to form a single, closed, Wilson loop (see Figure 1).

\begin{figure}[htb]
\begin{center}
\includegraphics[scale=0.7]{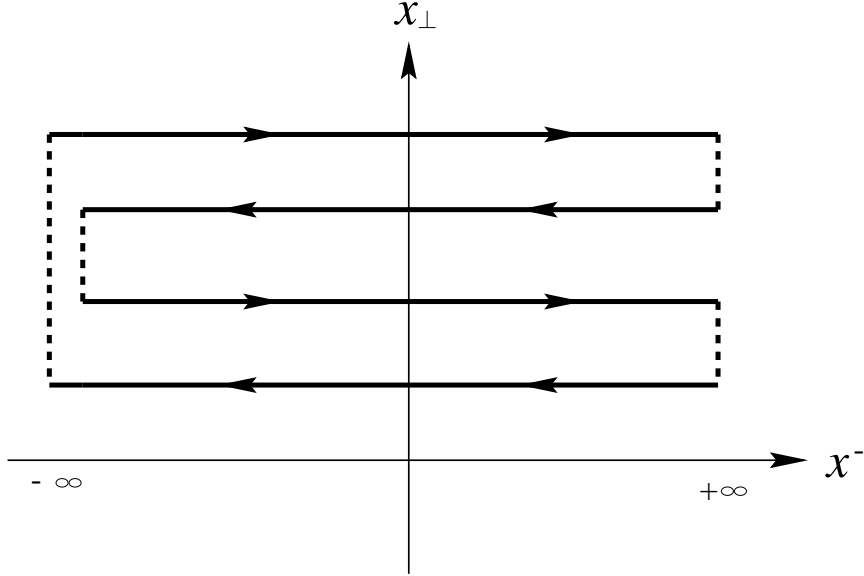}
    \caption{A closed Wilson loop for $\tr (V^\dag_x V_y V^\dag_z V_w )$
    as an example of the operator (\ref{gauge_inv_op}). Solid lines are
     Wilson lines along the longitudinal directions which are manifest
    on the operator, and
    dashed lines are Wilson lines which connect two end points
    at longitudinal infinity $x^-=\pm \infty$, but are in fact unity in
    the covariant gauge. }
\end{center}
\end{figure}

This observation suggests that there is an intimate relation
between the infrared finiteness and gauge invariance, that we would
like to clarify now. To this end, we first notice that the
differential operators which enter the finiteness conditions, namely
\BQA
{\mathcal G}_{\rm L}[\omega_{\rm L}]&\equiv&\, \omega^a_{\rm L}
\int d^2 \x  \frac{\delta}{\delta \alpha^a_\tau(\x)}\, ,\\
{\mathcal G}_{\rm R}[\omega_{\rm R}]&\equiv&\, -\omega^a_{\rm R}
\int d^2\x  (\V_{\x})^{ab}
\frac{\delta}{\delta \alpha^b_\tau(\x)}\, ,
\EQA
are, respectively,
the generators of left and right, global, color rotations:
\BQA
V^\dag_{\x} \, &\to&\,  \Omega_{\rm L}\, V^\dag_{\x}, \label{left}\\
V^\dag_{\x} \, &\to&\,   V^\dag_{\x}\, \Omega_{\rm R}^\dag ,\label{right}
\EQA
where $\Omega_{\rm L/R}=\exp (ig \omega_{\rm L/R}^a t^a)$
is a constant SU($N_c$) matrix. Indeed,
by using the formulae analogous to
Eq.~(\ref{formulae}), one can explicitly check that
the infinitesimal color rotations are given by
\BQA
\delta_{\rm L} V^\dag_{\x} &=&{\mathcal G}_{\rm L}[\omega_{\rm L}]V^\dag_{\x} =
ig \omega_{\rm L} V^\dag_{\x},\\
\delta_{\rm R} V^\dag_{\x} &=& {\mathcal G}_{\rm R}[\omega_{\rm R}]V^\dag_{\x}=
- ig V^\dag_{\x} \omega_{\rm R}.
\EQA
Thus, for any operator ${\mathcal O}$ built with Wilson lines,
the finiteness conditions (\ref{finite_cond1}) and
(\ref{finite_cond2}) are tantamount to the conditions of
invariance under such global color rotations. It is
trivial to check that the operator (\ref{gauge_inv_op}) is indeed
invariant under the color rotations (\ref{left}) and (\ref{right}).

Now, with their one--sided action, the color rotations (\ref{left})
and (\ref{right}) look {\it a priori} different from the ordinary gauge
transformations of the Wilson lines. Still, as we explain now, they are
in fact the {\it residual gauge transformations} with respect to the
`covariant gauge' in which the CGC theory is usually formulated (see
\cite{CGCreviews} for details).
We recall here that, in this gauge, the vector potential has only one
non--zero component, the light--cone field $A^+_a\equiv \alpha_a$, which
is independent of $x^+$. This structure of the field is preserved
by a gauge transformation
$A^\mu\to \Omega (A^\mu + \frac{i}{g}\del^\mu)\Omega^\dag $ in which
the gauge function $\Omega$ depends only upon $x^-$. Thus a residual
gauge transformation amounts to
\BQA
A^+ \to \Omega(x^-) \left(A^+ + \frac{i}{g}\del^+\right)\Omega^\dag(x^-),\quad
\Omega(x^-)={\rm e}^{ig \omega^a(x^-)t^a}, \label{gauge_A+}
\EQA
which induces the following transformation for the Wilson line $V^\dag_{\x}$ :
\BQA
V^\dag_{\x} \ \to\ \Omega(x^-=\infty)\, V^\dag_{\x} \, \Omega^\dag(x^-=-\infty).
\EQA
So far, the gauge function $\omega(x^-)$ is arbitrary. The "global"
color rotations introduced in Eqs.~(\ref{left}) and
(\ref{right}) are now obtained as the two (independent) special gauge
transformations in which $\omega(x^-)\to 0$ (and thus $\Omega(x^-)\to 1$)
at either $x^-\to -\infty$ (for the left rotation
(\ref{left})) or $x^-\to +\infty$ (for the right rotation (\ref{right})).
This establishes the interpretation of the finiteness conditions as the
conditions for gauge symmetry\footnote{Note also that the left and right generators
form two independent SU($N_c$) algebra, as is expected for gauge
transformations. More precisely, the differential operators
$
J_{L}^a(\x) \equiv -\frac{1}{ig} \frac{\delta}{\delta \alpha_{\x}^a},
$
and
$J_{R}^a(\x) \equiv \frac{1}{ig} (\V_{\x})^{ab}
           \frac{\delta}{\delta \alpha_{\x}^b}$,
which enter the definitions of the left and right generators,
satisfy the following algebra [$\delta_{xy}\equiv \delta^{(2)}(\x-\y)$]
\BQA
 [J_{L}^a(\x),\, J_{L}^b(\y)] = if^{abc}J_{L}^c(\x) \delta_{xy},\quad
 [J_{R}^a(\x),\, J_{R}^b(\y)] = if^{abc}J_{R}^c(\x) \delta_{xy},\nonumber
\EQA
and commute with each other ${}[J_{L}^a(\x), J_{R}^b(\y)]=0.$}.

Finally, let us show the infinitesimal transformation
for the gauge field integrated over the longitudinal direction:
$\alpha^a(\x)=\int dx^- \alpha^a(x^-,\x)$. As we will discuss shortly,
this is a natural variable in the weak--field regime.
Since $A^+=\alpha(x^-,\x)$
transforms as in Eq.~(\ref{gauge_A+}), its infinitesimal
change is given by $\delta \alpha(x^-,\x)=
\del^+\omega(x^-)$. Therefore, $\alpha^a(\x)$ transforms as
\BQA
\alpha^a(\x) \longrightarrow  \ \alpha^a(\x)
+\int_{-\infty}^\infty dx^- \del^+ \omega^a(x^-)
\ = \ \alpha^a(\x) + \xi^a, \label{gauge_alpha}
\EQA
with $\xi^a\equiv \omega^a(x^-=\infty)-\omega^a(x^-=-\infty)$ being a
pure number.
This transformation law will be useful in checking the gauge
invariance of various operators in the weak--field limit.

\setcounter{equation}{0}
\section{Weak--field regime and Green's functions in the CGC}
\label{WEAK_FIELD}

In order to make contact with previous evolution equations proposed in
perturbative QCD at high energy, like the BFKL \cite{bfkl} and BKP
\cite{bartels,pra} equations, which neglect saturation effects and
therefore apply only in the dilute regime at relatively high transverse
momenta (well above the saturation scale), it is convenient to consider
also the weak--field limit of the JIMWLK equation (here, with dipolar
kernel). When $g\alpha \ll 1$, one can expand the Wilson lines in
Eq.~(\ref{eta}) to lowest nontrivial order, to obtain
 \be\label{expV} (1- \tilde
 V^\dagger_{z} \tilde
 V_{x})^{fa}\,\approx\,ig\big(\alpha^c({\x}) -
 \alpha^c({\z})\big) \big(T^c \big)_{fa},\ee
where
 \be\label{alphaT} \alpha^a(\x)\,\equiv\,\int
 dx^-\,\alpha^a(x^-,\x)\,\equiv\,\alpha^a_{\x}\ee
 is the effective color field in the transverse plane,
as obtained after integrating over the longitudinal profile of the
hadron. Then, the kernel $\eta$ becomes simply quadratic in  $
\alpha^a_{\x}$, and  the evolution equation
(\ref{JIMWLK_evolution_simple}) reduces to \BQA \frac{\del}{\del
\tau}\langle {\mathcal O}\rangle_\tau
&=&\frac{g^2}{16\pi^3}\int_{xyz}\frac{(\x-\y)^2}{(\x-\z)^2(\z-\y)^2}\NN
&&\qquad \times \left\langle
(\alpha_{\x}-\alpha_{\z})^a(\alpha_{\y}-\alpha_{\z})^b f^{acf} f^{bfd}
\frac{\delta}{\delta\alpha^c_\tau(\x)}\frac{\delta}
{\delta\alpha_\tau^d(\y)}{\mathcal O} \right\rangle_\tau\!.\,\,
\label{weak} \EQA
 For consistency with the above manipulations, the observable
${\mathcal O}$ --- which in general is some operator built with
Wilson lines (see, e.g., Eq.~(\ref{gauge_inv_op})) --- must be
correspondingly expanded in powers of $g\alpha$. Note that the
relevant order for this expansion depends upon the operator at hand,
and also upon the channel that we consider for scattering (see Sect.
\ref{OPERATORS}). For instance, for the dipole operator shown in
Eq.~(\ref{dipole}), the lowest nontrivial contribution is obtained
after expanding the Wilson lines up to {\it second} order in
$g\alpha$ :
 \be \label{Vexp} \hspace*{-0.5cm}
 V^\dagger_{{x}}[\alpha] &\approx
 &1\,+\, {i}g\int dx^-
\alpha^a(x^-,{\x}) t^a\\
&-&\frac{g^2}{2} \int \! dx^-\!\int\! dy^-
\alpha^a(x^-,{\x})\alpha^b(y^-,{\x})\big[\theta(x^- - y^-)t^a t^b +
\theta(y^- - x^-)t^b t^a\big].\nonumber \ee
 Clearly, to this order, the ordering of the color matrices
$\alpha^a(x^-,{\x})t^a$ in $x^-$ starts to play a role in the
expansion of the Wilson lines. But this ordering is still irrelevant
for the computation of the dipole $S$--matrix to lowest order,
because of the symmetry of the color trace in Eq.~(\ref{dipole}) :
${\rm tr} (t^a t^b)=\frac{1}{2}\,\delta^{ab}={\rm tr} (t^b t^a)$.
Namely, one finds: \BQA S(\x,\y;\alpha)
 \equiv \frac{1}{N_c} \tr (V_{\x}^\dag V_{\y})\,\simeq\,
1-\frac{g^2}{4N_c}(\alpha_{\x}^a-\alpha_{\y}^a)^2,\label{nnn} \EQA
 where the final expression involves only the two--dimensional
field $\alpha_{\x}^a$, cf. Eq.~(\ref{alphaT}). The latter property
turns out to be shared by all the operators for weak (single)
scattering that we shall discuss throughout this work. This feature,
together with the structure of the simplified evolution Hamiltonian
manifest on Eq.~(\ref{weak}), enable us to consider the weak field
approximation to the CGC effective theory as being restricted to the
Hilbert space of the functions built with $\alpha_{\x}^a$. Onto this
space, the functional derivatives in Eq.~(\ref{weak}) can be trivially
replaced by ${\delta}/{\delta\alpha^a_{\x}}$. Thus, in this
approximation, the {\it longitudinal structure} of the color field
becomes irrelevant.

Note that the weak--field Hamiltonian in Eq.~(\ref{weak}) is quadratic
both in $\alpha$ and in the functional derivative with respect to
$\alpha$. Thus, when acting on the $n$--point function $\langle
\alpha(x_1)\alpha(x_2)\cdots \alpha(x_n)\rangle_\tau$, this Hamiltonian
does not change the number $n$ of fields. This has an important
consequence for the weak--field evolution described by
Eq.~(\ref{weak}): During this evolution, the number of gluons in the
$t$-channel remains {\it fixed}, familiar to the `multi--reggeons'
approaches based on BFKL evolution
\cite{bartels0,bartels,pra,lipatov,ko}, to which we shall eventually
compare the present formalism. In particular, this implies that the
JIMWLK evolution in the weak--field regime is {\it linear}.

 It is already known \cite{jklw97,iancu} that, in the weak field
limit, the JIMWLK evolution of the dipole $S$--matrix reduces to the
corresponding BFKL equation \cite{bfkl}. This can be easily checked on
the first Balitsky equation, Eq.~(\ref{Balitsky_fdt}), by first
rewriting this equation in terms of the {\it dipole scattering
amplitude},
 \be\label{Pomeron}
 -iT(\x,\y;\alpha)\equiv 1 -  S(\x,\y;\alpha)
 = 1 -\frac{1}{N_c} \tr (V_{\x}^\dag V_{\y})\,,\ee
and then linearizing the ensuing equation with respect to $T$,
which is formally\footnote{We ignore here the subtleties
associated with fluctuations in the dilute regime which in general
render such a linearization illegitimate even when $|\langle
T\rangle_\tau|\ll 1$; see Ref. \cite{it04} for details.}
appropriate in the weak scattering regime where $|T| \ll 1$.  But
for the following developments in this paper it is still
instructive to give a rapid derivation of the BFKL equation, by
using directly the weak field approximation in Eqs.~(\ref{weak})
and (\ref{nnn}). Specifically, in this approximation $-i\langle
T(\x,\y) \rangle_\tau \simeq \langle N(\x,\y) \rangle_\tau$, with:
 \BQA \langle N(\x,\y) \rangle_\tau &\simeq& \frac{g^2}{4N_c}\langle
 (\alpha_{\x}^a-\alpha_{\y}^a)^2\rangle_\tau\NN
 &\equiv &\frac{g^2}{4N_c}\,\left[
 f_\tau(\x,\x)+f_\tau(\y,\y)-2f_\tau(\x,\y)\right],
 \label{NF} \EQA
with the following definition for the 2--point Green's function of
the color fields in the dilute regime:
 \be\label{f2def}
  f_\tau(\x,\y) \equiv \langle\alpha^a(\x)\alpha^a(\y)\rangle_\tau
   = f_\tau(\y,\x).
 \ee
Note that, although colorless (it carries no open color indices), the
object in Eq.~(\ref{f2def}) is still not gauge invariant: the residual
gauge transformation for $\alpha^a_{\x}$ consists in the constant shift
$\alpha^a_{\x} \to \alpha^a_{\x} + \xi^a$ [cf.
Eq.~(\ref{gauge_alpha})], but $f_\tau(\x,\y)$ is not invariant under
this operation. On the other hand, the linear combination yielding the
scattering amplitude (\ref{NF}) is clearly invariant, as it should
(since Eq.~(\ref{NF}) has been obtained after a consistent expansion of
the gauge--invariant operator (\ref{Pomeron})). Thus, {\it a priori}, one is
allowed to use the dipolar form of the evolution equation,
Eq.~(\ref{weak}), for the physical amplitude $\langle
(\alpha_{\x}^a-\alpha_{\y}^a)^2\rangle_\tau$, but not also for the
Green's function  $f_\tau(\x,\y)$. Still, since Eq.~(\ref{weak}) is
linear in ${\mathcal O}$, it is clear that the equation obeyed by
$\langle (\alpha_{\x}^a-\alpha_{\y}^a)^2\rangle_\tau$ is correctly
obtained by separately evolving each of the Green's functions in the
second line of Eq.~(\ref{NF}), and then summing up the corresponding
results. This argument shows that, in fact, it is legitimate to use the
dipolar evolution equation (\ref{weak}) even for quantities which by
themselves are {\it not} gauge invariant, so like the Green's function
(\ref{f2def}), provided these quantities are eventually used as
building blocks in the construction of gauge--invariant observables. In
such a case, the use of the dipolar Hamiltonian $H_{{\rm dp}}$ can be
seen as a convenient prescription to regulate the infrared
singularities which would appear at intermediate steps when using the
original JIMWLK Hamiltonian in Eqs.~(\ref{JIMWLK})--(\ref{eta}).

Specifically, by using Eq.~(\ref{weak}) for ${\mathcal
O}=\alpha^a_{\x}\alpha_{\y}^a$, one easily finds
 \BQA\label{eqf2} \frac{\partial}{\partial \tau}
 f_\tau(\x,\y)&=&
  \frac{\bar{\alpha_s}}{2\pi} \int d^2\z
 \frac{(\x-\y)^2}{(\x-\z)^2(\y-\z)^2}\\
 &&\ \times \Big(
 f_\tau(\x,\z)+f_\tau(\y,\z)-f_\tau(\x,\y)-f_\tau(\z,\z)
 \Big).\nonumber
 \EQA
This equation is well defined both in the infrared (because of the
rapid decay of the dipole kernel at large values of $\z$), and in the
ultraviolet (the short--distance poles of the kernel at $\x=\z$ and
$\y=\z$ are actually harmless because they have zero residue). By
contrast, the equation which is obtained by acting on
$\alpha^a_{\x}\alpha_{\y}^a$ with the (weak--field version of the)
original JIMWLK Hamiltonian\footnote{This equation can be found, e.g.,
as Eq.~(3.10) in Ref. \cite{IM04}.} contains terms which are
potentially singular at large distances.

By using Eqs.~(\ref{NF}) and (\ref{eqf2}), one finds the coordinate
space (or `dipolar') version of the BFKL equation for $\langle
N(\x,\y) \rangle_\tau$, as expected:
 \BQA \frac{\partial}{\partial \tau}
\langle N(\x,\y)\rangle_\tau&=&\frac{\bar{\alpha}_s}{2\pi} \int
d^2\z
\frac{(\x-\y)^2}{(\x-\z)^2(\y-\z)^2} \label{bfkl}\\
&&\qquad \times \Big(\langle N(\x,\z)\rangle_\tau+\langle
N(\z,\y)\rangle_\tau -\langle N(\x,\y)\rangle_\tau\Big). \nonumber
 \EQA
In this last equation, ultraviolet finiteness is ensured by the
vanishing of the scattering amplitude at equal points, $\langle
N(\x,\x)\rangle=0$, a property consistent with Eq.~(\ref{NF}) and
which reflects `color transparency'.

In what follows, we shall take Eq.~(\ref{eqf2}) (supplemented with an
appropriate initial condition) as the {\it definition} of the 2--point
Green's function in the CGC formalism. This situation illustrates a
general feature of the present approach, namely the fact that, because
of infrared complications, the definition of Green's functions meets
with ambiguities which disappear only in the construction of
gauge--invariant quantities. Of course, these Green's functions are
only {\it intermediate} objects, which are strictly speaking not
needed: it is always possible to construct directly the equation obeyed
by the gauge--invariant quantity of interest, which is then free of any
ambiguity. The reason why it is nevertheless convenient to introduce
such Green's functions, it is because these are the CGC analogs of the
multi--gluon exchanges considered in the more traditional approaches to
high energy QCD \cite{bartels0,bartels,pra,lipatov,ko}, to which we
would like to establish a connection in the forthcoming sections.

\setcounter{equation}{0}
\section{Odderon operators in the CGC}
\label{OPERATORS}

With this section, we start our discussion of the odderon within the
CGC framework. To start with, we shall construct the operators
describing multiple odderon exchanges in the scattering between the CGC
and a relatively simple projectile, such as a color dipole, or three
quarks in a colorless state. The dipole--CGC scattering can be viewed
as a sub--process of the diffractive scattering of a virtual photon on
some dense hadronic target (the CGC), a process in which odderon
contributions are expected, e.g., in the production of $C$--even mesons
like $\eta_c$ (see, e.g., \cite{BLV,bartels2}). As for the
3--quark system, this may be viewed as a crude `valence quark' model of
the baryon.

\subsection{The dipole-CGC scattering}

Consider first the simplest case, namely the high energy scattering of
a $q\bar q$ dipole off the CGC, and let us briefly outline the
construction of the corresponding $S$--matrix. As explained in Sect.
\ref{review}, we need to first compute the $S$--matrix for a {\it
fixed} configuration of the classical field $\alpha$, and then average
over the latter. For the first step, we can use the eikonal
approximation: $S(\x,\y;\alpha)=\langle {\rm out} \vert {\rm in}
\rangle $, where the transverse positions of the quark ($\x$) and the
antiquark ($\y$) are the same in the in--coming and the out--going
states. One can write, schematically, ($i=1,\cdots,N_c$ is the color
index) \BQA \ket{\rm \, in}\sim \bar\psi_i^{\rm in}(\x)\psi_i^{\rm
in}(\y)\ket{0},\qquad \ket{\rm \, out}\sim \bar\psi_i^{\rm
out}(\x)\psi_i^{\rm out}(\y)\ket{0},\nonumber \EQA where the
appropriate normalization is understood. The relation between the
in--coming and the out--going fields is found by solving (the
high--energy version of) the Dirac equation $(\del_- - ig \alpha^a t^a
)\psi=0$ for a given gauge field configuration $\alpha$. This implies
that $ \psi^{out}_i= (V^\dag_{\x})_{ij} \psi^{in}_j$ with the Wilson
line in the fundamental representation. The $S$-matrix becomes:
 \BQA S(\x,\y;\alpha) =\langle
{\rm out} \vert {\rm in} \rangle
&=&\frac{1}{N_c}(V_{\x}^\dag)^{ij}(V_{\y})^{ki}\delta^{kl}\delta^{jl}
 =\frac{1}{N_c}\tr (V^\dag_{\x} V_{\y}),\nonumber \EQA
(we have restored the appropriate normalization), in agreement with
Eq.~(\ref{dipole}). The physical $S$--matrix is finally obtained after
averaging over the random classical color field, cf.
Eq.~(\ref{averaging}), an operation which also introduces the
dependence upon the energy (i.e., upon $\tau$): \BQA\label{Sdipole}
S_\tau(\x,\y)=\frac{1}{N_c}\langle \tr (V^\dag_{\x} V_{\y})
 \rangle_\tau. \EQA
Since non--linear in $\alpha$, the above formula describes in general
{\it multiple} exchanges, which can be either even, or odd, under the
operation of charge conjugation $C$. To single out $C$--even
(`pomeron') or $C$--odd (`odderon') exchanges, one needs to project
Eq.~(\ref{Sdipole}) onto incoming and outgoing states with appropriate
$C$--parities. Since the charge conjugation for fermions is defined by
$ C \psi C^{-1}= -i (\bar\psi \gamma^0\gamma^2 )^T,$ and $C \bar \psi
C^{-1}= (-i\gamma^0\gamma^2 \psi)^T,$ it is clear that the eigenstates
of $C$ in the dipole sector are given by $(\bar\psi(\x)\psi(\y)\pm
\bar\psi(\y)\psi(\x))\ket{0}$, where $+$($-$) sign yields the
$C$-even(odd) state. These structures are natural because the charge
conjugation is essentially the exchange of a quark and an antiquark.

Taking the $C$--odd dipole state as the in--coming state (this is the
state selected by the virtual photon wavefunction), and the $C$--even
dipole state as the out--going state (this would be selected by a
$\eta_c$ meson), one obtains the following $C$--odd contribution to the
$S$-matrix:
 \BQA\label{Sodd}
 S^{\rm odd}(\x,\y) =\langle\, {\rm out,\, even}\, | \,{\rm in, \,
odd}\, \rangle = \frac{1}{2N_c} \, \left\langle \tr(V^\dag_{\x} V_{\y})
 - \tr(V^\dag_{\y} V_{\x})\right\rangle_\tau. \EQA
This allows us to identify the operator for $C$--odd exchanges in the
dipole--CGC scattering ("the dipole odderon operator") as
 \BQA O(\x,\y)
\equiv \frac{1}{2iN_c}\tr
(V_{\x}^{\dagger}V_{\y}-V_{\y}^{\dagger}V_{\x})\,=\,-\, O(\y,\x),
\label{dipole_odderon}
 \EQA
where the factor of $i$ is introduced in order for this quantity to be
real: indeed, since $V$ and  $V^{\dagger}$ are unitary matrices, we
have $[\tr (V_{\x}^{\dagger}V_{\y})]^*= \tr (V_{\y}^{\dagger}V_{\x})$.

One can directly check, by using the transformation property of the
gauge fields under charge conjugation, \BQA \label{CA} C\, A_\mu \,
C^{-1} = - (A_\mu)^T, \EQA that the operator (\ref{dipole_odderon}) is
indeed $C$--odd: For a generic Wilson line $V$ constructed with
$A_\mu$, Eq.~(\ref{CA}) implies \BQA \label{CV} C\, V \, C^{-1} =
(V^\dag)^T = V^*, \EQA so that $C \, \tr (V^\dag_{\x} V_{\y})\,
C^{-1}=\tr (V^\dag_{\y} V_{\x})$, or, finally, $C\, O(\x,\y) \, C^{-1}=
- O(\x,\y)$.

Note that the $C$--odd contribution (\ref{Sodd}) is the imaginary part
of the $S$--matrix element, or, equivalently, the real part of the
scattering amplitude $T$, with $S=1+iT$: \BQA \langle O(\x,\y)
\rangle_\tau =\Im {\rm m\,} S_{\tau}(\x,\y), \EQA which was to be
expected. Correspondingly,  the $C$--{even}, Pomeron exchange,
amplitude, that we shall denote as $N(\x,\y)$, is identified with the
real part of the $S$-matrix:
 \BQA
&&N(\x,\y)\equiv 1- \frac{1}{2N_c} \tr
(V_{\x}^{\dagger}V_{\y}+V_{\y}^{\dagger}V_{\x}),
\label{dipole_Pomeron}\\
&&\langle N(\x,\y)\rangle_\tau = 1-\Re {\rm e}\, S_\tau(\x,\y).
 \EQA
Operatorially, $S=1+iT = 1 - N + iO$. Note the obvious boundary
conditions which follow from $S_\tau(\x,\x)=1$ (or directly from the
definitions (\ref{dipole_odderon}), (\ref{dipole_Pomeron})):
 \BQA
 N(\x,\x)\,=\, O(\x,\x)\,=\,0. \label{boundary_cond} \EQA
Clearly, these conditions remain true after averaging over the random
field $\alpha$.

From perturbative QCD, we expect the lowest order contribution to the
odderon exchange to be represented by three gluons tied up together
with the $d^{abc}$ symbol, where $d^{abc}=2 \tr(\{t^a,t^b\}t^c)$ is the
totally symmetric tensor. The operator
$d^{abc}A_\mu^a(\x)A_\nu^b(\y)A_\rho^c(\z)$ is indeed $C$--odd, as
obvious from Eq.~(\ref{CA}). Let us check that a similar structure
emerges also from the CGC operator (\ref{dipole_odderon}) when this is
evaluated in the weak--field limit. The lowest non--trivial contribution
to Eq.~(\ref{dipole_odderon}) is obtained by expanding the Wilson lines
there up to {\it cubic} order in the field $\alpha$ in the exponent.
(The terms quadratic in $\alpha$ cancel in the difference of traces in
Eq.~(\ref{dipole_odderon}).) By collecting the remaining terms, one
obtains: \BQA O(\x,\y) \simeq \frac{-g^3}{24N_c}
  d^{abc}\left\{ 3(\alpha_{\x}^a\alpha_{\y}^b \alpha_{\y}^c-
\alpha_{\x}^a\alpha_{\x}^b \alpha_{\y}^c) +(\alpha_{\x}^a \alpha_{\x}^b
\alpha_{\x}^c-\alpha_{\y}^a \alpha_{\y}^b \alpha_{\y}^c)\right\}.
 \label{odd} \EQA
As expected, this expression is cubic in $\alpha^a$ with the color
indices contracted symmetrically by the $d$--symbol. Note that, because
of the symmetry properties of this symbol, the path--ordering of the
Wilson lines in $x^-$ has been irrelevant for computing $O(\x,\y)$.

The linear combination of trilinear field operators in Eq.~(\ref{odd})
is gauge invariant by construction. To render this more explicit, let
us rewrite this operator as
  \BQA O(\x,\y) \simeq \frac{-g^3}{24N_c}
  d^{abc}(\alpha_{\x}^a-\alpha_{\y}^a)(\alpha_{\x}^b- \alpha_{\y}^b)
 (\alpha_{\x}^c-\alpha_{\y}^c).\label{odd_another} \EQA
This is manifestly invariant under a residual gauge transformation,
which consists in a constant shift of the field: $\alpha^a \to \alpha^a
+ \xi^a$ (cf. Eq.~(\ref{gauge_alpha})). In fact, an alternative way to
construct the gauge--invariant linear combination appearing in
Eq.~(\ref{odd}) is to directly impose the (weak-field version of the)
finiteness conditions (\ref{finite_cond1}) and (\ref{finite_cond2}) on
the $C$--odd Green's function $d^{abc}\langle
\alpha^a_{\x}\alpha^b_{\y} \alpha^c_{\y} \rangle_\tau$.

\subsection{The 3-quark--CGC scattering}

To describe the 3--quark colorless state, we shall use the following
"baryonic" operator\footnote{A similar approach was adopted for the
proton-proton scattering in Refs.~\cite{do,dosch}.} $\epsilon^{ijk}
\psi^i(\x)\psi^j(\y)\psi^k(\z)$, where $\epsilon^{ijk}$ is the complete
antisymmetric symbol, and the color indices $i,j,k$ can take the values
1,2, or 3. Thus, the construction below applies only for $N_c=3$. (The
generalization to arbitrary $N_c$ is in principle possible, but the
analysis becomes more complicated because the baryonic operator is then
built with $N_c$ quark fields.) By using the same technique as for
dipole--CGC scattering, one obtains the following $S$--matrix: \BQA
S_\tau(\x,\y,\z)=\frac{1}{3!}\,\epsilon^{ijk}\epsilon^{lmn} \langle
V^{\dagger}_{il}(\x)V^{\dagger}_{jm}(\y)V^{\dagger}_{kn}(\z)
\rangle_{\tau}, \EQA where $\x,\,\y,$ and $\z$ are the transverse
positions of the three quarks. This operator is symmetric under any
permutations of the three coordinates, and is normalized as
$S_\tau(\x,\x,\x)=1.$ By using Eq.~(\ref{CA}), it is easy to check that
the odderon contribution is given again by the imaginary part of the
$S$-matrix :
 \BQA\label{S3q} \langle B(\x,\y,\z)\rangle_\tau = \Im {\rm
m} \, S_\tau(\x,\y,\z), \EQA where the "3--quark odderon operator"
$B(\x,\y,\z)$ has been defined as\BQA
B(\x,\y,\z)=\frac{1}{3!2i}\left(\epsilon^{ijk}\epsilon^{lmn}
V^{\dagger}_{il}(\x)V^{\dagger}_{jm}(\y)V^{\dagger}_{kn}(\z)- {\rm
c.c.}\right).\label{3q_odderon}
 \EQA
This is totally symmetric too, and satisfies the boundary condition
$B(\x,\x,\x)=0$, which is an immediate consequence of the normalization
$S_\tau(\x,\x,\x)=1$.

The simplest way to see that the operator (\ref{3q_odderon}) is gauge
invariant is to notice that this can be rewritten in terms of the
manifestly gauge invariant operators shown in Eq.~(\ref{gauge_inv_op}).
Indeed, by using the identity\footnote{We have already used a similar
relation when we have specified the normalization of
$S_\tau(\x,\y,\z)$.} \BQA \frac{1}{3!}\epsilon^{ijk}\epsilon^{lmn}
V_{il}(\w)V_{jm}(\w)V_{kn}(\w)=\det V(\w)=1, \EQA where $\w$ is an
arbitrary transverse coordinate, one can equivalently rewrite the
3--quark odderon operator $B$  as \BQA \hspace*{-0.8cm}
 B(\x,\y,\z)
&=&\frac{1}{3!2i}\Bigl[\, \tr
 (V_{\x}^\dag V_{\w})\tr
(V_{\y}^\dag V_{\w}) \tr (V_{\z}^\dag V_{\w}) -\tr(V_{\x}^\dag V_{\w})
\tr(V_{\y}^\dag V_{\w} V_{\z}^\dag V_{\w})\NN &&\quad  -\, \tr (V_{\y}^\dag
V_{\w}) \tr (V_{\x}^\dag V_{\w} V_{\z}^\dag V_{\w})
  -\tr (V_{\z}^\dag V_{\w}) \tr (V_{\x}^\dag V_{\w} V_{\y}^\dag V_{\w})\NN
&&\quad +\, \tr (V_{\x}^\dag V_{\w} V_{\y}^\dag V_{\w} V_{\z}^\dag V_{\w})
  + \tr( V_{\x}^\dag V_{\w} V_{\z}^\dag V_{\w} V_{\y}^\dag V_{\w}) -{\rm c.c.}\,
  \Bigr].
 \label{3q_odderon1} \EQA
Note that this expression involves not only dipolar operators, but also
higher multi-polar ones (quadrupoles and sextupoles). By construction,
this expression is independent off $\w$ when $N_c=3$. Thus, it can be
simplified by choosing $\w$ to be one of the quark coordinates, say
$\w=\z$. Then, Eq.~(\ref{3q_odderon1}) reduces to:
 \BQA B(\x,\y,\z) \,=\, \frac{1}{3!2i}\Bigl [ \tr
(V_{\x}^\dag V_{\z})\tr (V_{\y}^\dag V_{\z})
  -\tr (V_{\x}^\dag V_{\z} V_{\y}^\dag V_{\z})- {\rm c.c.}\Bigr].\label{3q_odderon2}
 \EQA
which looks indeed considerably simpler, but where the symmetry in
$\x,\,\y$ and $\z$ is not manifest anymore (although, for $N_c=3$, we
know that this expression {\it is} totally symmetric, by construction).

In particular, when two of the coordinates are the same, the 3--quark
odderon operator reduces to the dipole odderon operator,
Eq.~(\ref{dipole_odderon}):
 \BQA
 B(\x,\z,\z)\,=\,O(\x,\z)\,=\,-B(\x,\x,\z). \qquad (N_c=3) \label{avo} \EQA
This is physically reasonable, because the diquark state is equivalent
to an antiquark as far as color degrees of freedom are concerned, and
can be most easily checked by setting $\y=\z$ in the r.h.s. of
Eq.~(\ref{3q_odderon2}). Note however that, if one rather sets $\y=\x$
in the same expression, then it is not immediately obvious that the
ensuing expression for $B(\x,\x,\z)$ is indeed equal to $-O(\x,\z)$ when
$N_c=3$, as it should. This is so because of the lack of manifest
symmetry of Eq.~(\ref{3q_odderon2}), as alluded to above. Still, by
diagonalizing  the unitary matrix $V_x^\dag V_z$, and after performing
some simple algebraic manipulations (relying on the fact that the
corresponding eigenvalues $\lambda_i$, $i=1,2,3$, are pure phases and
obey $\lambda_1\lambda_2\lambda_3=1$), it is possible to check that the
equality $B(\x,\x,\z) = -O(\x,\z) =-B(\x,\z,\z)$ holds indeed.

In the weak field approximation, as obtained after expanding to lowest
non--trivial order (i.e., to cubic order in $\alpha$) the Wilson lines
in any of the previous expressions for $B(\x,\y,\z)$, one finds again a
gauge invariant linear combination of trilinear field operators with
the color indices contracted with the $d$--symbol:
 \BQA\label{sy}
&&\hspace*{-1cm} B(\x,\y,\z)\simeq
\frac{g^3}{144}d^{abc}\\
&&\hspace*{-6mm}\times\left\{
(\alpha_{\x}^a-\alpha_{\z}^a)+(\alpha_{\y}^a-\alpha_{\z}^a)\right\}
\left\{
(\alpha_{\y}^b-\alpha_{\x}^b)+(\alpha_{\z}^b-\alpha_{\x}^b)\right\}
\left\{
(\alpha_{\z}^c-\alpha_{\y}^c)+(\alpha_{\x}^c-\alpha_{\y}^c)\right\}.
\nonumber  \EQA
In fact, in this weak--field regime, the 3--quark $C$--odd operator
is fully determined by gauge symmetry together with the requirement
of total symmetry with respect to the external coordinates: Indeed,
it can be checked that Eq.~(\ref{sy}) uniquely emerges from the
$C$--odd Green's function (\ref{f3def}) after symmetrization and
imposing the finiteness condition (\ref{finite_cond1}).

\setcounter{equation}{0}
\section{Odderon evolution in the dipole--CGC scattering}
\label{ODD_DIPOLE}

In this and the next sections, we shall apply the general JIMWLK
equation (in its dipolar form, cf. Eq.~(\ref{JIMWLK_evolution_simple}))
to the operators describing odderon exchanges constructed in the
previous section, in order to deduce the evolution equations for the
respective, $C$--odd, scattering amplitudes.

We start with the simpler case of the dipole scattering, for which we
shall discuss separately the weak--field limit (corresponding to a
single scattering), and the general, non--linear, case (which includes
multiple scattering). In fact, in this particular case, it is rather
straightforward to write down directly the non--linear equations (see
below),
from which the equations for the weak--field limit can be then simply
deduced by linearization. Still, the more detailed approach that we
shall follow below is instructive as a preparation for the more tedious
case of the 3--quark operator, to be discussed in Sect. \ref{ODD_3Q}.

\subsection{Linear evolution and the odderon Green's function}

In the weak--field regime, where we can limit ourselves to a single
odderon exchange, it is convenient to proceed as in Sect.
\ref{WEAK_FIELD} and introduce the (totally symmetric) {\it odderon
Green's function}
 \BQA f_\tau(\x,\y,\z)\equiv d^{abc}\langle
 \alpha_{\x}^a \alpha_{\y}^b
 \alpha_{\z}^c\rangle_\tau\, ,\label{f3def}
 \EQA
in terms of which the weak--field version of the dipole odderon
operator (\ref{odd}) can be rewritten in a form similar to
Eq.~(\ref{NF}) for the pomeron:
 \BQA \hspace*{-0.5cm} O(\x,\y) \simeq \frac{-g^3}{24N_c}
   \Big\{ 3\Big(f_\tau(\x,\y,\y) -  f_\tau(\x,\x,\y)\Big)
   + f_\tau(\x,\x,\x) - f_\tau(\y,\y,\y)
 \Big\}.\NN \label{odd_green}
 \EQA
The discussion of the pomeron Green's function (\ref{f2def}) in Sect.
\ref{WEAK_FIELD} applies to the odderon Green's function (\ref{f3def})
as well: The latter is not a gauge--invariant quantity, so its
evolution under the original JIMWLK Hamiltonian would be afflicted by
infrared singularities, which can however be regulated by using the
(weak field version of the) dipolar Hamiltonian $H_{{\rm dp}}$. This
yields a mathematically well defined equation for $f_\tau(\x,\y,\z)$,
that we shall use as the {\it definition} of the odderon Green's
function.

Still as in the pomeron case, the equation obeyed by the scattering
amplitude (\ref{odd_green}) --- which {\it is} gauge--invariant --- is
not sensitive to the ambiguities which affect the definition of the
Green's function, and comes out the same whatever form of the
Hamiltonian is used in its derivation. In fact, the only reason for
introducing the Green's function (\ref{f3def}) is the fact that it is
for this function that we shall verify the BKP equation later on.

Specifically, by using Eq.~(\ref{weak}) for ${\mathcal
O}=d^{abc}\alpha^a_{\x}\alpha_{\y}^b\alpha_{\z}^c$, and after some
lengthy algebra, one obtains
 \be
\frac{\partial}{\partial \tau} f_\tau(\x,\y,\z)\,\, &=&\,\,
  \frac{\bar{\alpha_s}}{4\pi} \int d^2\w
\frac{(\x-\y)^2}{(\x-\w)^2(\y-\w)^2} \NN
&&\Big(
f_\tau(\x,\w,\z)+f_\tau(\w,\y,\z)
-f_\tau(\x,\y,\z)
 -f_\tau(\w,\w,\z) \Big)\NN
&&\quad +\,\,\Big\{{\rm 2\  cyclic\  permutations}\Big\}\,,
 \label{eqf3} \ee
which is like applying the equation (\ref{eqf2}) for the 2--point
Green's function to each pair of points within $f_\tau(\x,\y,\z)$. It
can be checked as before that this equation is well defined both in the
infrared and in the ultraviolet. It is now straightforward to construct
the equation obeyed by the linear combination in Eq.~(\ref{odd_green}),
and thus find that this is precisely the BFKL equation (\ref{bfkl})
 \BQA
\frac{\partial}{\partial \tau} \langle O(\x,\y)\rangle_\tau
&=&\frac{\bar{\alpha}_s}{2\pi} \int d^2\z
\frac{(\x-\y)^2}{(\x-\z)^2(\y-\z)^2} \NN &&\times \Big(\langle
O(\x,\z)\rangle_\tau +\langle O(\z,\y)\rangle_\tau -\langle
 O(\x,\y)\rangle_\tau \Big) \label{bfkl_odd} \EQA
(we have also used the identity
$f_{eag}f_{dbg}d_{abc}=\frac{N_c}{2}d_{edc}$), in agreement with an
original observation by Kovchegov, Szymanowski and Wallon \cite{kov}.
In view of the formal difference between the pomeron and the odderon
operators, as given by Eqs.~(\ref{NF}) and,  respectively, (\ref{odd}),
it may appear as a surprise that they obey both the same evolution
equation. But this becomes more natural if one remembers that $N$ and
$O$ are, respectively, the imaginary part and the real part of the
scattering amplitude $T$ (with $S=1+iT$), and that in the weak
field regime $T$ obeys a linear equation with real coefficients
--- the linearized version of the first Balitsky equation,
Eq.~(\ref{Balitsky_fdt}) --- which is therefore separately satisfied
by its real and imaginary parts.

But since the initial conditions corresponding to $N$ and $O$ are
different (in particular, they have different $C$--parities), so are
also the respective solutions, and their behaviors at high energy. In
Appendix A, we compute within the CGC formalism the $C$--odd initial
conditions for some simple targets: a bare quark and a $q\bar q$
dipole. Namely, if the target is a single quark with transverse
position $\x_0$, one obtains
  \BQA \langle
 O(\x,\y)\rangle_{\tau=0}=\frac{\alpha_s^3}{12}
 \frac{(N_c^2-4)(N_c^2-1)}{N_c^3} \ln^3\frac{|\x-\x_0|}{|\y-\x_0|},
 \label{initial_quark} \EQA
whereas for the more interesting case of a dipolar target (with the
quark being at $\x_0$ and the antiquark at $\y_0$), one rather finds
 \BQA \langle O(\x,\y)\rangle_{\tau=0}=\frac{\alpha_s^3}{12}
\frac{(N_c^2-4)(N_c^2-1)}{N_c^3}
\ln^3\frac{|\x-\x_0||\y-\y_0|}{|\x-\y_0||\y-\x_0|}.
\label{initial_dipole} \EQA These expressions are in agreement with the
corresponding results in Ref.~\cite{kov} up to an overall numerical
factor. As expected, these initial conditions are antisymmetric under
the exchange of $\x$ and $\y$, and thus satisfy the boundary condition
(\ref{boundary_cond}). It is easily checked that this boundary
condition is preserved by the evolution according to
Eq.~(\ref{bfkl_odd}), as necessary for this equation to be well
defined. 

The high--energy behavior of the odderon solution to
Eq.~(\ref{bfkl_odd}) has been analyzed too in Ref.~\cite{kov}, where it
has been shown that the projection of the general BFKL solution onto
$C$--odd initial conditions selects ($C$--odd) BFKL eigenfunctions
whose maximal intercept is equal to one. It turns out that these are
the same eigenfunctions that were previously identified,  by Bartels,
Lipatov, and Vacca \cite{BLV}, as exact solutions to the
momentum--space BKP equation \cite{bartels,pra}. Thus, in contrast to
the pomeron solution to the BFKL equation, which at high energy rises
exponentially with $Y$ ($\,N(Y)\sim {\rm e}^{(\alpha_\mathbb P -1) Y}$,
with the BFKL intercept $\alpha_\mathbb P = 1+ (4\ln 2)\alpha_s
N_c/\pi$ \cite{bfkl}), the corresponding odderon solution rises only
slowly, as a power of $Y\sim \ln s$.

But, of course, these types of high--energy behavior (for either $N$ or
$O$), which are mathematical consequences of the BFKL equation, are
physically acceptable only so far as this equation is a correct
approximation, that is, within the limited range of energies where the
unitarity corrections are indeed negligible. For higher energies, the
evolution is governed by more complicated, non--linear, equations, to
which we now turn.

\subsection{Non--Linear evolution}

For dipole--CGC scattering, the general evolution equations obeyed by
the average amplitudes $\langle N(\x,\y) \rangle_\tau$ and $\langle
O(\x,\y)\rangle_\tau$ in the strong field regime can be easily inferred
from the first Balitsky equation (\ref{Balitsky_fdt}) : Since the
operators $N(\x,\y)$ and $O(\x,\y)$ are, respectively, the real part
and the imaginary part of the dipole $S$--matrix $S(\x,\y) =
(1/N_c)\tr(V_{\x}^{\dagger}V_{\y})$, cf. Eqs.~(\ref{dipole_odderon})
and (\ref{dipole_Pomeron}), it is clear that the respective equations
can be simply obtained by separating the real part and the imaginary
part in Eq.~(\ref{Balitsky_fdt}). One thus obtain:
 \BQA \frac{\partial}{\partial
\tau} \langle O(\x,\y)\rangle_\tau
 &=&\frac{\bar\alpha_s}{2\pi}\int
d^2\z \frac{(\x-\y)^2}{(\x-\z)^2(\z-\y)^2} \NN &&\quad  \times\,
\Big\langle O(\x,\z)+ O(\z,\y)- O(\x,\y) \NN && \qquad -\,
O(\x,\z)N(\z,\y)-
N(\x,\z) O(\z,\y)\Big\rangle_\tau,\ \ \label{11} \\
\frac{\del}{\del \tau}\langle N(\x,\y) \rangle_\tau
&=&\frac{\bar\alpha_s}{2\pi}\int d^2\z
\frac{(\x-\y)^2}{(\x-\z)^2(\z-\y)^2}\NN &&\quad \times\, \Big\langle
N(\x,\z)+ N(\z,\y)- N(\x,\y)\NN &&\qquad -\,  N(\x,\z)N(\z,\y) +\,
 O(\x,\z) O(\z,\y) \Big\rangle_\tau. \label{22} \EQA
As is generally the case for the Balitsky equations, the equations
above do not close by themselves, but rather belong to an infinite
hierarchy. Interestingly, the non--linear terms in these equations
couple the evolution of $C$--odd and $C$--even operators. For instance,
the last term, quadratic in $O$, in the r.h.s. of Eq.~(\ref{22}) for
$\langle N\rangle_\tau$ describes the merging of two odderons into one
pomeron. This process has not been recognized in previous studies of
the Balitsky hierarchy, but the vertex connecting one pomeron to two
odderons has been already computed in lowest order perturbation theory
\cite{BE99}, and it would be interesting to compare such previous
results with the corresponding vertex in Eq.~(\ref{22}) (which is
essentially the dipole kernel).

But the odderon--pomeron coupling which turns out to have the most
dramatic consequences is the one encoded in the last terms in
Eq.~(\ref{11}) for $\langle O\rangle_\tau$ : As we shall shortly argue,
this coupling leads to a rather rapid suppression of the $C$--odd
contributions to scattering in the high energy regime where unitarity
corrections start to be important (i.e., where $\langle N\rangle_\tau
\sim {\mathcal O}(1)$). To construct the argument without having to
resort to an infinite hierarchy of equations, we shall restrict
ourselves to the mean field approximation in which the non--linear
terms in Eqs.~(\ref{11})--(\ref{22}) are assumed to factorize. In the
strong field regime, which will be our main focus below, we expect this
approximation to be qualitatively correct \cite{IMFLUCT}.

In this mean field approximation, Eqs.~(\ref{11})--(\ref{22}) reduce
to a closed system of coupled, non--linear, equations for $\langle N
\rangle_\tau$ and $\langle O \rangle_\tau$ :
 \BQA \hspace*{-0.7cm} \frac{\partial}{\partial \tau}
\langle O(\x,\y)\rangle_\tau &=&\frac{\bar\alpha_s}{2\pi}\int d^2\z
\frac{(\x-\y)^2}{(\x-\z)^2(\z-\y)^2} \label{evolution_odd_fact}\\
&& \quad\times\Bigl [ \langle O(\x,\z)\rangle_\tau + \langle O(\z,\y)
\rangle_\tau - \langle O(\x,\y)\rangle_\tau \NN && \qquad -\, \langle
O(\x,\z)\rangle_\tau \langle N(\z,\y)\rangle_\tau - \langle
N(\x,\z)\rangle_\tau \langle O(\z,\y)\rangle_\tau\Bigr ], \ \nonumber
\EQA
 \BQA \hspace*{-0.7cm} \frac{\del}{\del \tau}\langle
N(\x,\y) \rangle_\tau &=&\frac{\bar\alpha_s}{2\pi}\int d^2\z
\frac{(\x-\y)^2}{(\x-\z)^2(\z-\y)^2}\label{evolution_even_fact}\\
&& \quad \times \Bigl[ \langle  N(\x,\z)\rangle_\tau + \langle
N(\z,\y)\rangle_\tau - \langle N(\x,\y)\rangle_\tau \NN && \qquad -\,
\langle N(\x,\z) \rangle_\tau \langle N(\z,\y) \rangle_\tau +\, \langle
O(\x,\z)\rangle_\tau \langle O(\z,\y)\rangle_\tau\Bigr]. \nonumber
 \EQA
The first of these equations has been already proposed in
Ref.~\cite{kov}, as a plausible non--linear generalization of
Eq.~(\ref{bfkl_odd}). As for Eq.~(\ref{evolution_even_fact}), this is
the Kovchegov equation \cite{K} supplemented by a new term describing
the merging of two odderons.

The Kovchegov equation has been extensively studied over the last few
years, both analytically and numerically, and although the exact
solution is not known, its general properties are by now well
understood
\cite{LT99,AB01,Motyka,LL01,GBS03,Nestor03,Nestor04,SCALING,MT02,DT02,MP03}.
In its most synthetic description, due to Munier and Peschansky
\cite{MP03} (see also \cite{AB01}), this solution can be viewed as a
front connecting the saturation regime where\footnote{Until the end of
this section, we shall mostly use the simplified notations $N$ and $O$
for the respective expectation values.} $N=1$
--- this is reached for dipole sizes $r=|\x-\y|$ much larger than
the saturation length $1/ Q_s(\tau)$ --- to the unstable regime at
$r\ll 1/ Q_s(\tau)$, where the amplitude is small ($N\ll 1$), but it
rises rapidly with $\tau$, according to BFKL equation. When increasing
$\tau$, the front propagates towards smaller values of $r$ (or larger
transverse momenta), and its instantaneous position defines the
saturation momentum $Q_s(\tau)$. The latter is found to rise
exponentially with $\tau$: $Q_s^2(\tau)=Q_0^2\, {\rm e}^{c\bar\alpha_s
\tau}$, with $c$ a numerical constant determined by the BFKL dynamics
\cite{GLR,SCALING,MT02,DT02}.

As we shall argue below, this mean--field picture of the pomeron
exchanges is not significantly modified by the odderon contribution to
Eq.~(\ref{evolution_even_fact}). In particular, the values $N=1$ and
$N=0$ remain as (stable and, respectively, unstable) fixed points of
the evolution described by
Eqs.~(\ref{evolution_odd_fact})--(\ref{evolution_even_fact}), but to
them one should add a new fixed point, namely $O=0$, which is
asymptotically approached at high energy.

It is first easy to check that $N=1$ and $O=0$ are indeed fixed points
at high energy. To also see that this is the {\it only} combination of
fixed points in this limit, notice from Eq.~(\ref{evolution_odd_fact})
that, when increasing $N$, the non--linear terms in this equation act
towards suppressing the odderon \cite{kov}. It is thus consistent to
assume that, at high energy, the odderon contribution represents only a
small perturbation to the Kovchegov equation, so that the pomeron
saturates in the standard way: $N(r,\tau)\simeq 1$ for $r\gg
1/Q_s(\tau)$. Also, the dominant energy behavior of the saturation
momentum, i.e., the value of the saturation exponent $c$, should not
change, since this is fully determined by the linear (BFKL) part of the
equation for $N$. Using this assumption, it is possible to study the
approach of $O$ towards zero, and thus check the consistency of our
hypothesis.

Namely, for sufficiently high energy, such that $r\gg 1/Q_s(\tau)$,
the integral in the r.h.s. of Eq.~(\ref{evolution_odd_fact}) is
dominated by relatively large dipoles, for which $N\simeq 1$. In
this regime, the non--linear terms in this equation precisely cancel
the first two linear terms there, and the equation simplifies to
 \BQA \hspace*{-0.8cm}
 \frac{\partial }{\partial \tau}\langle O(\x,\y)\rangle_\tau
\simeq - \bar{\alpha}_s\!\int_{Q_s^{-2}(\tau)}^{r^2}\!{dz^2\over z^2}\,
\langle
 O(\x,\y)\rangle_\tau= - {\bar{\alpha}_s} \ln[Q_s^2(\tau)r^2]\,\langle
 O(\x,\y)\rangle_\tau, \EQA
which together with $\ln[Q_s^2(\tau)r^2]=c\bar{\alpha}_s(\tau-\tau_0)$
immediately implies:
  \BQA \hspace*{-0.3cm}\langle O(\x,\y)\rangle_\tau \simeq \exp
\left\{-\frac{c}{2}\,\bar{\alpha}_s^2(\tau-\tau_0)^2\right\} \langle
O(\x,\y)\rangle_{\tau_0}\quad {\rm for}\quad r\gg 1/Q_s(\tau).
 \label{Oddasymp} \EQA
As anticipated, this is a rapidly decreasing function of $\tau$,
which is actually the same as the function describing the approach
of the real part of the $S$--matrix towards the `black--disk'
limit $S=0$ (the Levin-Tuchin law) \cite{LT99,SAT}. We expect the
fluctuations neglected in the mean field approximation leading to
Eqs.~(\ref{evolution_odd_fact})--(\ref{evolution_even_fact}) to
modify (actually, decrease) the value of the overall coefficient
in the exponent of Eq.~(\ref{Oddasymp}), but preserve the above
qualitative picture \cite{IMFLUCT}.

\setcounter{equation}{0}
\section{Odderon evolution in the scattering of the 3--quark system}
\label{ODD_3Q}

The new feature which makes the 3--quark system conceptually
interesting is the fact that the corresponding scattering amplitude
depends upon three independent transverse coordinates. Therefore,
already the lowest--order odderon amplitude, as given in
Eq.~(\ref{sy}), involves configurations in which the three exchanged
gluons are attached to different quark legs, and which are thus
probing the complete functional dependence of the odderon Green's
function defined in Eq.~(\ref{f3def}). As we shall further argue in
Sect. \ref{Sect_BKP}, this in turns implies that the 3--quark---CGC
scattering is a good theoretical laboratory to study the general
solution to the BKP equation.

Our discussion in this section will be restricted to the weak--field
version of the 3--quark odderon operator, Eq.~(\ref{sy}), which
describes scattering via the exchange of a single odderon. This is
indeed sufficient to discuss the correspondence with the BKP equation
in the next section. Within the CGC formalism, there is no difficulty
of principle (other than the tediousness of the corresponding algebraic
manipulations) which would prevent us from deriving the evolution
equations satisfied by the general, non--linear, 3--quark operators in
Eqs.~(\ref{3q_odderon}) or (\ref{3q_odderon1}). However, these general
equations are complicated and not very illuminating: Through their
non--linear terms, they couple the 3--quark operator to other color
structures. This is especially manifest if one uses the form
(\ref{3q_odderon1}) of the 3--quark amplitude: this involves operators
of various multi-polar orders, which in the Balitsky hierarchy are
coupled to other operators of even higher multi-polar moments.

By using the expression of $B(\x,\y,\z)$ as a linear combination of
odderon Green's functions, as manifest in Eq.~(\ref{sy}), together with
the equation (\ref{eqf3}) obeyed by the latter, one can deduce after a
straightforward but lengthy calculation the following linear evolution
equation for $\langle B_{\x\y\z}\rangle_\tau \equiv \langle
B(\x,\y,\z)\rangle_\tau$:
 \BQA
\frac{\partial}{\partial \tau} \langle B_{\x\y\z}\rangle_\tau
&=&\frac{3\alpha_s}{4\pi^2} \int d^2\w
\frac{(\x-\y)^2}{(\x-\w)^2(\y-\w)^2} \NN &&\qquad \times \Bigl( \langle
B_{\x\w\z}\rangle_\tau + \langle B_{\w\y\z}\rangle_\tau -\langle
B_{\x\y\z} \rangle_\tau\NN && \qquad\quad - \langle
B_{\w\w\z}\rangle_\tau -\langle B_{\x\x\w}\rangle_\tau-\langle
B_{\y\y\w}\rangle_\tau-\langle B_{\x\y\w}\rangle_\tau \Bigr)\NN &&+\,
({\rm 2\ cyclic\ permutations}). \label{prot}
 \EQA
(Of course, the same equation could be obtained also directly from
Eq.~(\ref{weak}) with ${\mathcal O}= B_{\x\y\z}$, but the corresponding
manipulations would be even more tedious; the introduction of the
Green's function (\ref{f3def}) at intermediate steps has also the
advantage to better organize the calculation.) Note that
Eq.~(\ref{prot}) is a {\it closed} equation for $\langle
B_{\x\y\z}\rangle_\tau$, which was expected in view of gauge
invariance: the only gauge invariant operators available are $
B_{\x\y\z}$ and $O_{\x\y}=B_{\x\y\y}$ (cf. Eq.~(\ref{avo})).

Consider the structure of Eq.~(\ref{prot}) in some detail: The
linear combination of $B$'s in the integrand vanishes at the points
$\w=\x$ and $\w= \y$ where lie the poles of the dipole kernel, so
the latter are again innocuous. Also, one can easily check that the
above equation is consistent with the relation (\ref{avo}) between
the dipole and the 3--quark odderon amplitudes: if one sets $\z=\y$,
Eq.~(\ref{prot}) reduces indeed to Eq.~(\ref{bfkl_odd}) with
$N_c=3$. Lastly, Eq.~(\ref{prot}) manifestly preserves the symmetry
of the scattering amplitude under the permutation of its three
coordinate variables.

The initial condition can be calculated within the CGC formalism, in a
similar way as for dipole--CGC scattering (cf. Appendix A). For
example, for a single quark at transverse position $\x_0$, one finds
\BQA\label{proini}
\hspace*{-5mm}\langle B_{\x\y\z}\rangle_{\tau=0}\\
&&\hspace*{-14mm} =\frac{5}{3^4}\alpha_s^3 \ln
\frac{|\x-\x_0||\y-\x_0|}{|\z-\x_0|^2} \ln
\frac{|\y-\x_0||\z-\x_0|}{|\x-\x_0|^2} \ln
 \frac{|\z-\x_0||\x-\x_0|}{|\y-\x_0|^2}. \nonumber \EQA
As expected, this is symmetric under the permutation of the three
external coordinates, and satisfies the boundary condition $\langle
B_{\x\x\x} \rangle_\tau=0$.

In the next section, we shall argue that the high energy behavior of
the solution $\langle B_{\x\y\z}\rangle_\tau$ to Eq.~(\ref{prot}) is
controlled by the BLV solution \cite{BLV} to the BKP equation, and thus
has a maximal intercept equal to one.

\section{Comparison with previous approaches}
\label{Sect_BKP} \setcounter{equation}{0}

Within the traditional perturbative QCD approach to small--$x$
evolution, the odderon exchange is viewed as the exchange of a
composite object made of three reggeized gluons
which evolves with energy according to the
Bartels--Kwiecinski--Praszalowicz (BKP) equation \cite{bartels,pra}.
From the point of view of the CGC formalism, this exchange is a single
scattering, and should be compared to the weak--field limit of the
corresponding CGC equations, as derived in the previous sections. In
what follows, we shall demonstrate that the BKP odderon corresponds to
the CGC Green's function introduced in Eq.~(\ref{f3def}): The BKP
equation, which is traditionally written in momentum space, is
essentially the Fourier transform of Eq.~(\ref{f3def}) for
$f_\tau(\x,\y,\z)$. Based on our previous discussion of
gauge--invariant scattering amplitudes, we shall then argue that the
natural Hilbert space for discussing odderon exchanges is in fact
larger than the one which is generally used in the literature, and
within which the BFKL Hamiltonian shows holomorphic separability
\cite{lipatov,ko}.

\subsection{Relation to the BKP equation}

The BKP equation is traditionally written in momentum
space as
\BQA
\frac{\partial}{\partial \tau} F_\tau(\kk_1,\kk_2,\kk_3)
&=&\frac12 \sum_{i=1}^3\int d^2\kk_1'd^2\kk_2'd^2\kk_3'\,
     \delta^{(2)}(\kk_1'+\kk_2'+\kk_3'-\q) \, \delta^{(2)}(\kk_i -\kk_i ')\NN
&&\qquad \ \times H_{{\rm BFKL}}(\kk_{i-1},\kk_{i+1};\kk'_{i-1},\kk'_{i+1})
 F_\tau(\kk_1',\kk_2',\kk_3'),\label{BKP} \EQA
where $F_\tau(\kk_1,\kk_2,\kk_3)$ is the Green's function for the exchange
of three reggeized gluons (with transverse momenta $\kk_1$, $\kk_2$, and
$\kk_3$, respectively) in a $C$--odd, color singlet, state.
Furthermore, $\kk_4 \equiv \kk_1$ (and similarly $\kk_4'\equiv \kk_1'$), and
$\q$ is the momentum transfer. The factor $\frac12$ accounts  for the
fact that two of the gluons are in the color octet state. Finally,
$H_{{\rm BFKL}}$ is the non-forward BFKL kernel including the
virtual terms: \BQA
 H_{{\rm BFKL}}(\kk_1,\kk_2;\kk_1',\kk_2')
 &=&\frac{\kk_1^2 \kk_2'^2+\kk_2^2 \kk_1'^2-(\kk_1-\kk_1')^2(\kk_1+\kk_2)^2}
 {\kk_1^2 \kk_2^2(\kk_1-\kk_1')^2} \NN
 &&-\pi \delta^{(2)}(\kk_1'-\kk_1)\left(\ln \frac{\kk_1^2}{\epsilon^2} +\ln
 \frac{\kk_2^2}{\epsilon^2}\right),
 \label{nonforward}
 \EQA where $\epsilon$ is an infrared cutoff which does not affect
 physical results.

The function $F_\tau(\kk_1,\kk_2,\kk_3)$ is often referred to as a
`scattering amplitude', but in fact the physical amplitudes are
obtained only after convoluting this with appropriate impact factors. A
generic odderon amplitude reads:
 \BQA\label{TPhi}
{\mathcal O}(\q) &=& \int d^2\kk_1d^2\kk_2d^2\kk_3\,\delta^{(2)}
(\kk_1+\kk_2+\kk_3-\q)\NN &&\qquad \qquad \times\, \Phi_{{\rm proj}}(\kk_1,\kk_2,\kk_3)\,
 F_\tau(\kk_1,\kk_2,\kk_3) , \EQA
where gauge symmetry requires the projectile impact factor $\Phi_{{\rm
proj}}(\kk_1,\kk_2,\kk_3)$ to vanish when $\kk_i =0$ for some $i$
(indeed, a zero--momentum gluon `sees' the projectile as a whole, and
the latter is globally colorless). Note that, when writing the
scattering amplitude as in Eq.~(\ref{TPhi}), the impact factor of the
target is implicitly included in the definition of $F_\tau$. This is in
line with the general approach in this paper, where the target is a
generic `color glass condensate', and the CGC Green's functions like
those in Eqs.~(\ref{f2def}) or (\ref{f3def}) include the relevant
information about the target impact factor.

In the case where the projectile is a proton, the odderon impact factor is
expected in the form (see, e.g., \cite{dosch})
 \BQA
 \hspace*{-0.5cm}
\Phi_{{\rm proton}}(\kk_1,\kk_2,\kk_3)&=&\int d^2\x d^2\y d^2\z |\Psi_{\rm
proton}(\x,\y,\z)|^2 (2\, {\rm e}^{i\kk_1\x}-\, {\rm e}^{i\kk_1\y}-\, {\rm
e}^{i\kk_1\z})\NN &&\ \times (2\, {\rm e}^{i\kk_2\y}-\, {\rm e}^{i\kk_2\z}-\,
{\rm e}^{i\kk_2\x}) (2\, {\rm e}^{i\kk_3\z}-\, {\rm e}^{i\kk_3\x}-\, {\rm
 e}^{i\kk_3\y}), \label{impact_proton} \EQA
where $|\Psi_{\rm proton}(\x,\y,\z)|^2$ is the proton light--cone
wavefunction, with $\x$, $\y$, and $\z$ denoting the coordinates of
the three valence quarks relative to their barycenter. (The
condition $\x+\y+\z=0$ is implicit in the definition of $|\Psi_{\rm
proton}|^2$.) The exponential terms within the parentheses
correspond to all the possible attachments of the three exchanged
gluons to the quark lines in the proton. Note that $\Phi_{{\rm
proton}}(\kk_1=0,\kk_2,\kk_3)=0$, etc., as expected.

Furthermore, if the projectile is a virtual photon, the odderon
couples to the dipole component of the photon Fock space, so that:
 \BQA \Phi_{\gamma^*}(\kk_1,\kk_2,\kk_3)&=&\int dz d^2 \rr
|\Psi_{\gamma^*}(z,\rr)|^2\label{impact_photon}\\
&&\times (\, {\rm e}^{i\kk_1\frac{\rr}{2}}-\,
{\rm e}^{-i\kk_1\frac{\rr}{2}})(\, {\rm e}^{i\kk_2\frac{\rr}{2}}-
\, {\rm e}^{-i\kk_2\frac{\rr}{2}})
(\, {\rm e}^{i\kk_3\frac{\rr}{2}}-\, {\rm e}^{-i\kk_3\frac{\rr}{2}}),\nonumber
 \EQA
where $\rr$ is the dipole size, and $\Psi_{\gamma^*}$ is the
light--cone wavefunction describing the dissociation of the virtual
photon, and can be computed in perturbation theory with respect to
$\alpha_{\rm EM}$ \cite{niko}.

To make contact between this more traditional approach and our
previous results in this paper, it is convenient to introduce first
the momentum space version of the odderon Green's function
introduced in Eq.~(\ref{f3def}). We thus define:
 \BQA
f_\tau(\kk_1,\kk_2,\kk_3)\equiv \int \frac{d^2\x d^2\y d^2\z}{(2\pi)^6}\,
{\rm e}^{-i\kk_1\x-i\kk_2\y-i\kk_3\z} \, f_\tau(\x,\y,\z)\,
 ,\label{Fourier_Green} \EQA
in terms of which the scattering amplitudes $\langle O(\x,\y)
\rangle_\tau$ and $\langle B(\x,\y,\z)\rangle_\tau$ can be rewritten
as
 \BQA \langle O(\x,\y)\rangle_\tau
&=&\int d^2\kk_1d^2\kk_2d^2\kk_3
f_\tau(\kk_1,\kk_2,\kk_3)\label{ta}\\
&&\ \times(\, {\rm e}^{i\kk_1\x}-\, {\rm e}^{i\kk_1\y})(\, {\rm
e}^{i\kk_2\x}-\, {\rm e}^{i\kk_2\y})(\, {\rm e}^{i\kk_3\x}-\, {\rm
e}^{i\kk_3\y}) ,\nonumber \EQA
 and,  respectively,
\comment{
 \footnote{ {\bf By using the Bose symmetry $f_\tau(\kk_1,\kk_2,\kk_3)=f_\tau(\kk_2,\kk_1,\kk_3)$, etc., and the property
 $f_\tau(\kk_1,\kk_2,\kk_3)=f_\tau(-\kk_1,-\kk_2,-\kk_3)$ which follows from the structure of the BKP equation Eq.~(\ref{BKP}),
 one can show that
\BQA  \int d^2\bb \, \langle O(\x,\y)\rangle_\tau=0, \EQA
 where $\x=\bb+\rr/2$, $\y=\bb-\rr/2$, in accordance with the fact that
  the dipole odderon amplitude vanishes at zero momentum transfer.}} }
 \BQA
&&\hspace*{-1cm}\langle B(\x,\y,\z)\rangle_\tau\nonumber\\
&&=\int d^2\kk_1d^2\kk_2d^2\kk_3\, f_\tau(\kk_1,\kk_2,\kk_3) (2\, {\rm
e}^{i\kk_1\x}-\, {\rm e}^{i\kk_1\y}-\, {\rm e}^{i\kk_1\z})\NN &&\qquad
\qquad \times (2\, {\rm e}^{i\kk_2\y}-\, {\rm e}^{i\kk_2\z}-\, {\rm
e}^{i\kk_2\x}) (2\, {\rm e}^{i\kk_3\z}-\, {\rm e}^{i\kk_3\x}-\, {\rm
e}^{i\kk_3\y}).\label{four}\EQA

Note the similarity between the exponential factors in these
equations and those in Eqs.~(\ref{impact_proton}) and
(\ref{impact_photon}) for the impact factors. This reflects the fact
that the couplings between the exchanged gluons and the quarks (or
antiquarks) in the projectile have been explicitly included in our
definition of the scattering amplitudes. By inspection of the
previous equations, it becomes clear that the CGC Green's function
$f_\tau(\kk_1,\kk_2,\kk_3)$ should correspond to the function
$F_\tau(\kk_1,\kk_2,\kk_3)$ of the traditional BKP approach. To be able to
compare these quantities, one also needs the equation satisfied by
$f_\tau(\kk_1,\kk_2,\kk_3)$, which is obtained after taking a Fourier
transform in Eq.~(\ref{eqf3}). A lengthy but straightforward
calculation shows that the resulting equation is the same as the BKP
equation (\ref{BKP}) up to terms proportional to delta--functions
$\delta^{(2)}(\kk_i )\ (i=1,2,3)$, which are however irrelevant for the
calculation of the scattering amplitudes, since they do not
contribute to the convolutions in Eqs.~(\ref{TPhi}) or
(\ref{ta})--(\ref{four}). This demonstrates the equivalence between
the two formalisms, in so far as the single odderon exchanges are
considered.

\subsection{Comments on the Hilbert space for odderon solutions}
\label{Hilbert}

Following the remarkable discovery by Lipatov \cite{lipatov}  that the
BFKL Hamiltonian exhibits holomorphic separability when restricted to
functions which belong to the M\"obius space
--- by which we mean the functions $f_\tau(\x,\y,\z, \dots)$ which
vanish when any two coordinates coincide with each other
($f_\tau(\x,\x,\z,\dots)=0$, etc.) ---, much effort has been devoted
towards finding solutions to the BKP equation (and, more generally, to
its generalization which describes the exchange of $n$ reggeized
gluons) within this particular Hilbert space of functions. This
situation is mathematically appealing since the restriction of the BKP
equation to a given holomorphic sector describes a dynamical system
which has a sufficient number of hidden conserved charges (three in the
case of the odderon) to be completely integrable\footnote{For the
generalization of the BKP equation to a system of $n$ reggeized gluons,
with $n > 3$, the same property holds only in the large--$N_c$ limit in
which one can restrict the BFKL--like pairwise interactions to
neighboring gluons \cite{lipatov,lipatov1,ko}.} \cite{lipatov1,ko}, and
which in fact can be identified as the XXX Heisenberg model of spin
$s=0$ \cite{ko}.

However, as recently reiterated in Ref. \cite{bartels3}, a careful
inspection reveals that the full BKP Hamiltonian  contains extra
delta functions like $\delta^{(2)}(\x-\y)$ in addition to the
separable Hamiltonian given in \cite{lipatov,lipatov1,ko}. One can
discard these delta functions and restore holomorphic separability
by working in the M\"obius space. But this is only natural in the
case of the BFKL pomeron, i.e., for the 2--point Green's function
$f_\tau(\x,\y)$, since in that case one can always redefine
 \BQA
\tilde{f}_\tau(\x,\y)\equiv f_\tau(\x,\y)-\frac12 f_\tau(\x,\x)-\frac12
f_\tau(\y,\y), \label{subt}\EQA which ensures that
$\tilde{f}_\tau(\x,\x)=0$, without affecting the calculation of
scattering amplitudes\footnote{In fact, a brief comparison with
Eq.~(\ref{NF}) reveals that the subtracted 2--point Green's function
can be identified with the pomeron scattering amplitude.}: The
subtracted terms in Eq.~(\ref{subt}), being independent of one of the
coordinates, give vanishing contributions when convoluted with a
colorless impact factor. However, for the odderon problem $(n=3)$, the
restriction to the M\"obius space is a highly nontrivial issue --- it
cannot be simply achieved via subtractions which preserve the physical
amplitudes ---, and therefore has implications on the physical
relevance of the various solutions to the BKP equation.

And, indeed, among the two exact solutions to the (odderon) BKP
equations which do not vanish rapidly at high energies
\cite{janik,BLV}, only one of them
--- the Janik--Wosiek (JW) solution \cite{janik}, which has an
intercept slightly lower than one --- lies indeed in the M\"obius
representation, e.g.,
 \BQA f_\tau^{JW}(\x,\x,\z)=0,\EQA
and has been constructed by exploiting conformal symmetry and
integrability. But, clearly, this solution does not couple to a dipole
in the present, leading--logarithmic, approximation. Besides, although
in principle it can couple to a 3--quark system, the corresponding
amplitude constructed according to Eq.~(\ref{sy}) (where we identify
$d^{abc}\langle \alpha_{\x}^a \alpha_{\y}^b \alpha_{\z}^c\rangle_\tau
\equiv f_\tau^{JW}(\x,\y,\z)$) has the rather curious property to
vanish at equal points ($\langle B(\x,\z,\z)\rangle_\tau=0$, etc.), for
which there is no compelling physical justification. For instance, the
initial condition (\ref{proini}) does not have this property. Thus, the
solution to Eq.~(\ref{prot}) corresponding to this particular initial
condition will {\it not} belong to the M\"obius space.

Similarly, the other exact solution known for the odderon BKP equation,
due to Bartels, Lipatov, and Vacca (BLV) \cite{BLV}, does not belong to
the M\"obius space either. For this solution, the dominant intercept is
exactly one, so the corresponding amplitude has only a weak energy
dependence (weaker than any power of the energy). As explicitly
verified in Ref. \cite{kov}, using the BLV solution within
Eq.~(\ref{ta}) for the dipole odderon amplitude $\langle
O(\x,\y)\rangle_\tau$ is equivalent to constructing the general
$C$--odd solution to the BFKL equation (\ref{bfkl_odd}). Thus, the BLV
solution appears as the physical Green's function describing the
odderon exchange between a virtual photon and some generic target (in
the dilute regime where saturation effects in the target are
unimportant).

Furthermore, by inserting the BLV solution into Eq.~(\ref{four}), one
obtains a particular solution to Eq.~(\ref{prot}) for the amplitude
$\langle B(\x,\y,\z)\rangle_\tau$ describing odderon exchanges between
a 3--quark system and a dilute target\footnote{To fully determine the
physical amplitude, one must still convolute the BLV Green's function
with the target impact factor.}. The following argument suggests that
this particular solution should in fact yield the dominant behavior of
$\langle B\rangle_\tau$ at high energy; that is, without actually
solving Eq.~(\ref{prot}), one can conclude that the intercept for
$\langle B\rangle_\tau$ is exactly one, and is determined by the BLV
solution. The argument relies on the relation (\ref{avo}) between the
odderon amplitudes for the 3--quark system and that for the dipole. Let
$1+\omega$ denote the intercept for $\langle B\rangle_\tau$, that is,
  \BQA \langle
 B(\x,\y,\z)\rangle_\tau=h(\x,\y,\z;\tau)\,{\rm e}^{\omega \tau}, \EQA
where $h$ is only slowly varying with $\tau$ (slower than any
exponential). Setting $\y=\z$, one finds
 \BQA \langle
 O(\x,\z)\rangle_\tau=h(\x,\z,\z;\tau)\,{\rm e}^{\omega \tau}. \EQA
Since, on the other hand, $\langle O(\x,\z)\rangle_\tau$ must be a
$C$--odd solution to the BFKL equation (\ref{bfkl_odd}), the analysis
in Ref. \cite{kov} demonstrates that the largest possible value of
$\omega$ is $\omega=0$. This argument misses the component of the BKP
solution which vanishes at equal points. However, the largest intercept
of this kind, corresponding to the JW solution, is known to be less
than one \cite{janik}.

\section{Conclusions}

In this paper we have given the first discussion of the odderon problem
within the framework of the effective theory for the color glass condensate.
The genuinely CGC part of the analysis has been rather straightforward:
The operators describing $C$--odd exchanges between the
CGC and two simple projectiles ---a $q\bar q$ color dipole and a 3--quark
system --- have been constructed in the eikonal approximation, which includes
multiple scattering to all orders via Wilson lines. By acting with the JIMWLK
Hamiltonian on these operators, we have then deduced the evolution
equations satisfied by the corresponding scattering amplitudes. This
is a standard, but generally quite lengthy, procedure, that we have considerably
simplified by introducing the dipolar form of the JIMWLK Hamiltonian.
In particular, for a dipole projectile, we have recovered the equations
previously obtained in Ref. \cite{kov}, that we have generalized here
beyond the mean field approximation.

What turned out to be more subtle, however, was the correspondence
with the traditional perturbative QCD approach (in particular,
with the BKP equation) in the limit where the scattering is weak.
Although our equations become linear in this limit, they still apply
to gauge--invariant scattering amplitudes, and not directly to
Green's functions. Besides, they are {\it a priori} written in coordinate
space. But the use of the dipolar version of the JIMWLK Hamiltonian has allowed
us to write down a well--defined equation for the odderon Green's function,
which after Fourier transformation to momentum space turned out to be the same
as the BKP equation.

Our analysis has emphasized the subtlety involved in the Fourier
transformation of the BKP equation, and the importance of the structure
of the external probe for selecting physical solutions to this equation.
While the latter point was already stressed in Refs. \cite{BLV,kov} in the
context of the dipole scattering, the ability of our formalism to deal
with a 3--quark system (which has three independent coordinates)
has made this point even clearer, thus shedding light on the Hilbert
space to be used in relation with the BKP equation.
From the viewpoint of Eq.~(\ref{prot}), the solutions which belong
to the M\"obius representation (i.e., which vanish at equal points:
$\langle B(\x,\z,\z)\rangle_\tau=0$, etc.)
are very special ones, and are unlikely to be realized during the
evolution from physical initial conditions. Rather, the solutions
to Eq.~(\ref{prot}), and therefore also to the BKP equation, which are
singled out by our gauge--invariant 3--quark amplitude and the respective
initial conditions do {\it not} vanish at equal points.
This observation, together with the relation
Eq.~(\ref{avo}), has led us to conclude that the highest odderon
intercept for the 3-quark--CGC scattering is exactly one, so like
for the dipole--CGC scattering \cite{kov}, and is described again by the
Bartels--Lipatov--Vacca solution  \cite{BLV} to the BKP equation.

As mentioned in the Introduction, our weak field analysis
of the odderon problem ($n=3, \ C=-1$) is intended as a first step
in a systematic study of multireggeon ($n\ge 3$) exchanges in the CGC formalism,
with the aim of clarifying the relation between this formalism and
more traditional approaches, like GLLA. Although successful in
establishing the correspondence with the BKP equation, our previous
analysis has also revealed a few subtleties which may lead to difficulties
when trying to extend this approach to processes with more than three reggeons.
We have seen indeed that the Green's functions in the CGC formalism
make sense only as building blocks (in the sense of linear
combinations) for scattering amplitudes in the weak field limit. Thus,
in order to have a meaningful definition for a $n$--point Green's function
(corresponding to a $n$--reggeon exchange), one must first
identify appropriate gauge--invariant amplitudes whose weak--field expansion
starts at order $n$ in $\alpha^a(\x)$.
The potential difficulty with this approach, however, is that there is
{\it a priori} no guarantee that the nonlocality in $x^-$ inherent
in the Wilson lines will disappear in the expansion leading to the
Green's functions. Recall, for instance, our previous construction of
the weak--field odderon operators, Eqs.~(\ref{odd}) and (\ref{sy}), from the
corresponding non--linear operators, Eqs.~(\ref{dipole_odderon}) and (\ref{3q_odderon}),
respectively: In that case, the nonlocality in $x^-$ has disappeared from the final
results only `accidentally', because of the presence of the totally symmetric tensor
$d^{abc}$. An alternative procedure which looks promising would be to
construct the gauge--invariant linear combination directly in the weak--field
limit, starting with a non--invariant Green's function and imposing on it
the finiteness condition (\ref{finite_cond1}). It remains as an interesting
open question whether any of the methods mentioned above can be used to
construct {\it arbitrary} $n$--reggeon exchanges.  We leave this and related issues
for future work.

\vspace*{-0.8cm}
\section*{Acknowledgments}
\vspace*{-0.6cm}
We are grateful to Carlo Ewerz for useful discussions and for comments
on an early version of the draft.
 Y.H. thanks Dima Kharzeev, Anna Stasto and Kirill Tuchin for useful conversations.
  Two of the authors (K.I. and L.M.) thank Yukio Nemoto for the
discussion about the baryonic Wilson line operators.
We would like to thank Basarab Nicolescu for constantly
encouraging us to address the odderon problem in the
framework of the color glass condensate.
Y. H. is supported by Special Postdoctoral Research Program of RIKEN,
and K.I. is supported by the program, JSPS Postdoctoral Fellowships for
Research Abroad.
This manuscript has been authorized under Contract No. DE-AC02-98CH10886
with the U. S. Department of Energy.

\appendix
\section{$C$-odd initial conditions in the dipole-CGC scattering}

The gauge field $\alpha^a_{\x}$ is created by a color source $\rho^a$
in the target CGC. It is given by (in the covariant gauge) \BQA
\alpha^a_{\x} = \frac{1}{4\pi}\int d^2\z \ln \frac{1}{(\x-\z)^2
\mu^2}\, \rho^a(\z) , \EQA where $\mu$ is an infrared cutoff which will
disappear from the final results. For a single quark at transverse
position $\x_0$, the color source is given by \BQA \rho^a(\z)= Q^a
\delta^{(2)}(\z-\x_0), \EQA and for a dipole made of a quark at $\x_0$ and
an antiquark at $\y_0$, \BQA \rho^a(\z)=Q^a \left[\delta^{(2)}(\z-\x_0) -
\delta^{(2)}(\z-\y_0)\right], \EQA where $Q^a$ is the color charge $Q^a=g
\int \psi^\dag t^a \psi$. We obtain the initial conditions for the
evolution equation by substituting the gauge fields created by these
"unevolved" targets into Eq.~(\ref{odd}) and (\ref{odd_another}), and
taking the average over the "random" gauge field. For the unevolved
targets
 the average over the random configuration simply reduces to the color average.
 More precisely,
in order to evaluate the average $\langle Q^a Q^b Q^c\rangle$, we
replace the $c$-number charge $Q^a$ by a color matrix $gt^a$ and
take the trace $\frac{1}{N_c}\tr (gt^a\, gt^b\, gt^c )$. Thus, it
can be evaluated as follows: \BQA d^{abc}\langle Q^a Q^b Q^c\rangle\
&\longrightarrow& \ d^{abc} \frac{1}{N_c} \tr (gt^a\, gt^b\, gt^c )=
\frac{g^3}{N_c} d^{abc} \frac{1}{2}\tr (\{t^a,t^b\}t^c)\NN &&\quad
=\frac{g^3}{4N_c} d^{abc}d^{abc}=
\frac{g^3}{4N_c^2}(N_c^2-4)(N_c^2-1). \EQA This yields the results
(\ref{initial_quark}) and (\ref{initial_dipole}). The technique
adopted here for the average was first proposed by Iancu and Mueller
for the onium-onium scattering to show the equivalence between the
color dipole picture and the CGC formalism \cite{IM04}. In fact,
this procedure
 goes beyond the original McLerran-Venugopalan (MV) model. This is
because the MV model is formulated with the Gaussian random source and
the average of odd number of sources such as $\langle \rho^a \rho^b
\rho^c\rangle$ is simply vanishing. In order to describe the odderon,
it is necessary to extend the MV model so that there is nontrivial
correlation among three sources. On the other hand, such nontrivial
correlation is correctly encoded in the CGC framework, as we claim in
the present paper. Indeed, the dipole JIMWLK equation (\ref{weak}) in
the weak--field limit suggests a similar equation for the weight
function $W_\tau[\alpha]$, but the Gaussian weight function is not the
exact solution to this evolution equation. It arises only in the
mean-field like approximation discussed in Ref.~\cite{SAT}.


\begin{thebibliography}{999}

\bibitem{bfkl}E.~A.~Kuraev, L.~N.~Lipatov and V.~S.~Fadin,
Sov.\ Phys.\ JETP {\bf 44} (1976) 443; Sov.\ Phys.\ JETP {\bf 45}
(1977) 199; I.~I.~Balitsky and L.~N.~Lipatov,
Sov.\ J.\ Nucl.\ Phys.\  {\bf 28} (1978) 822.

\bibitem{lipatov2} L.~N.~Lipatov,
Sov.\ Phys.\ JETP {\bf 63} (1986) 904.

\bibitem{nico} L.~Lukaszuk and B.~Nicolescu,
Lett.\ Nuovo Cim.\  {\bf 8} (1973) 405; D.~Joynson, E.~Leader,
B.~Nicolescu and C.~Lopez,
Nuovo Cim.\ A {\bf 30} (1975) 345.



\bibitem{bartels}J.~Bartels,
Nucl.\ Phys.\ B {\bf 175} (1980) 365.

\bibitem{pra}J.~Kwiecinski and M.~Praszalowicz,
Phys.\ Lett.\ B {\bf 94} (1980) 413.

\bibitem{Jaro}T.~Jaroszewicz, Acta. Phys. Polon. B{\bf 11} (1980) 965.

\bibitem{bartels0} J.~Bartels, Nucl.\ Phys.\ B {\bf 151} (1979) 293.




\bibitem{lipatov} L.~N.~Lipatov,
Phys.\ Lett.\ B {\bf 251} (1990) 284;  see also L.~N.~Lipatov,
``{\it Pomeron in Quantum Chromodynamics}",  in {\it Perturbative
Quantum Chromodynamics}, Ed. A.~H.~Mueller, World Scientific,
Singapore, 1989.

\bibitem{lipatov1} L.~N.~Lipatov, Phys.\ Lett.\ B {\bf 309}
(1993) 394;  JETP Lett.\  {\bf 59} (1994) 596; ``{\it High-energy
asymptotics of multicolor QCD and exactly solvable lattice
 models},'' hep-th/9311037.

\bibitem{ko} L.~D.~Faddeev and G.~P.~Korchemsky,
Phys.\ Lett.\ B {\bf 342} (1995) 311; G.~P.~Korchemsky,
Nucl.\ Phys.\ B {\bf 443} (1995) 255.




\bibitem{janik}R.~A.~Janik and J.~Wosiek,
Phys.\ Rev.\ Lett.\  {\bf 82}, (1999) 1092; J.~Wosiek and
R.~A.~Janik,
Phys.\ Rev.\ Lett.\  {\bf 79}, (1997) 2935.

\bibitem{BLV} J.~Bartels, L.~N.~Lipatov and G.~P.~Vacca,
Phys.\ Lett.\ B {\bf 477}, (2000) 178.



\bibitem{ko2}G.~P.~Korchemsky, J.~Kotanski and A.~N.~Manashov,
Phys.\ Rev.\ Lett.\  {\bf 88} (2002) 122002; S.~E.~Derkachov,
G.~P.~Korchemsky and A.~N.~Manashov,
Nucl.\ Phys.\ B {\bf 617} (2001) 375; S.~E.~Derkachov,
G.~P.~Korchemsky, J.~Kotanski and A.~N.~Manashov,
Nucl.\ Phys.\ B {\bf 645} (2002) 237; S.~E.~Derkachov,
G.~P.~Korchemsky and A.~N.~Manashov,
Nucl.\ Phys.\ B {\bf 661} (2003) 533.

\bibitem{de} H.~J.~De Vega and L.~N.~Lipatov,
Phys.\ Rev.\ D {\bf 64} (2001) 114019; Phys.\ Rev.\ D {\bf 66} (2002)
074013.

\bibitem{ewerz} C.~Ewerz,
{\it`` The odderon in quantum chromodynamics''}, hep-ph/0306137.

\bibitem{it04}
E.~Iancu and D. N. Triantafyllopoulos, ``{\it A Langevin equation for high
energy evolution with pomeron loops},'' hep-ph/0411405.


\bibitem{GLR}   L.V.~Gribov, E.M.~Levin, and M.G.~Ryskin, { Phys.
Rept. } {\bf 100} (1983) 1.

\bibitem{MQ86}  A.H.~Mueller and J.~Qiu, { Nucl. Phys.} {\bf
B268} (1986)  427.


\bibitem{MV94}
L. McLerran and R.~Venugopalan, { Phys.\ Rev.}\ {\bf D49} (1994)
2233; {\it ibid.} {\bf 49} (1994) 3352; {\it ibid.} {\bf 50} (1994)
2225.

\bibitem{mueller}A.~H.~Mueller,
Nucl.\ Phys.\ B {\bf 415} (1994) 373; Nucl.\ Phys.\ B {\bf 437} (1995)
107.


\bibitem{MP94}
A. H. Mueller and B. Patel, { Nucl. Phys.} {\bf B425} (1994) 471.


\bibitem{bartels1}
J.~Bartels, DESY 91-074 (unpublished); Phys.\ Lett.\ B {\bf 298} (1993)
204; Z. Phys. C {\bf 60} (1993) 471.

\bibitem{BW95}
J.~Bartels and M. W\"usthoff, Z. Phys. C {\bf 66} (1995) 157.

\bibitem{BV99}
M. Braun and G.P. Vacca, Eur. Phys. J. C {\bf 6} (1999) 147.

\bibitem{braun}
M.~Braun, Eur. Phys. J. C6 (1999) 321. 

\bibitem{BE99} J.~Bartels and C.~Ewerz,
JHEP {\bf 9909} (1999) 026; J.~Bartels, M.~Braun and G.~P.~Vacca,
``{\it Pomeron vertices in perturbative QCD in diffractive scattering},''
 hep-ph/0412218.

\bibitem{ewerz0} C.~Ewerz, Phys.\ Lett.\ B {\bf 472} (2000) 135;
{\it ibid.} {\bf 512} (2001) 239;  C.~Ewerz and V. Schatz, Nucl.\
Phys.\ A {\bf 736} (2004) 371.

\bibitem{bartels3}J.~Bartels, L.N.~Lipatov and G.P.~Vacca,
Nucl. Phys. {\bf B706} (2005) 391.

\bibitem{BE05}
C.~Ewerz and S. Braunewell, ``{\it The C-Odd Four-Gluon State in the
Color Glass Condensate},'' hep-ph/0501110.



\bibitem{RP97} R.~Peschanski, Phys.\ Lett.\ B {\bf 409} (1997) 491.





\bibitem{CGCreviews}
E.~Iancu, A.~Leonidov and L.~McLerran, in {\it QCD Perspectives on Hot
and Dense Matter}, Eds. J.-P. Blaizot and E. Iancu,  Kluwer Academic
Publishers (2002), hep-ph/0202270; \\
A. H. Mueller, in {\it QCD Perspectives on Hot and Dense Matter}, Eds.
J.-P. Blaizot and E. Iancu,  Kluwer Academic
Publishers (2002), hep-ph/0111244; \\
E.~Iancu and R. Venugopalan, in {\it Quark-Gluon Plasma 3}, Eds. R. C.
Hwa and X.-N. Wang, World Scientific (2003), hep-ph/0303204.







\bibitem{jklw97} J.~Jalilian-Marian, A.~Kovner, A.~Leonidov and H.~Weigert,
Nucl.\ Phys.\ B {\bf 504} (1997) 415; Phys.\ Rev.\ D {\bf 59}
(1999) 014014.

\bibitem{iancu} E.~Iancu, A.~Leonidov and L.~D.~McLerran,
Nucl.\ Phys.\ A {\bf 692} (2001) 583;
{ Phys. Lett.} {\bf B510} (2001) 133; E.~Ferreiro, E.~Iancu,
A.~Leonidov and L.~McLerran,
Nucl.\ Phys.\ A {\bf 703} (2002) 489;

\bibitem{weigert}
H.~Weigert,
Nucl.\ Phys.\ A {\bf 703} (2002) 823.


\bibitem{path}
J.~P.~Blaizot, E.~Iancu and H.~Weigert,
Nucl.\ Phys.\ A {\bf 713} (2003) 441.

\bibitem{balitsky}
I.~Balitsky, { Nucl.\ Phys.}\ {\bf B463} (1996) 99; { Phys. Rev.
Lett.} {\bf 81} (1998) 2024; { Phys. Lett.} {\bf B518} (2001) 235;
{\it High-energy QCD and Wilson lines}, hep-ph/0101042.

\bibitem{SAT} E.~Iancu and L.~McLerran, Phys. Lett. B{\bf 510} (2001) 145;
 E.~Iancu, K.~Itakura and L.~McLerran,
Nucl.\ Phys.\ A {\bf 724} (2003) 181.


\bibitem{MSW05}
A.H.~Mueller, A. Shoshi, and S. Wong, ``{\it Extension of the
JIMWLK equation in the low gluon density region},'' hep-ph/0501088.



\bibitem{K}  Yu. V. Kovchegov, { Phys. Rev.}
{\bf D60} (1999), 034008; {\it ibid.} {\bf D61} (2000) 074018.



\bibitem{IMFLUCT} E.~Iancu and A.H.~Mueller, Nucl. Phys.
A{\bf 730} (2004) 494.


\bibitem{MS04}
A.H. Mueller and A.I. Shoshi, { Nucl.\ Phys.}\  {\bf B692} (2004)
175.


\bibitem{IMM04}
E.~Iancu, A.H. Mueller, and S. Munier, {\it ``Universal behavior of QCD
amplitudes at high energy from general tools of statistical physics"},
hep-ph/0410018.


\bibitem{kov}Y.~V.~Kovchegov, L.~Szymanowski and S.~Wallon,
Phys.\ Lett.\ B {\bf 586} (2004) 267.



\bibitem{for}J.~R.~Forshaw and D.~A.~Ross,
``{\it Quantum Chromodynamics and the Pomeron},''  Cambridge University
Press, Cambridge, 1997.




\bibitem{IM04}E.~Iancu and A.H.~Mueller, Nucl. Phys.
A{\bf 730} (2004) 460.



\bibitem{bartels2}  J.~Bartels, M.~A.~Braun,
D.~Colferai and G.~P.~Vacca,
Eur.\ Phys.\ J.\ C {\bf 20} (2001) 323.


\bibitem{do}H.G.~Dosch, E.~Ferreira and A.~Kramer,
Phys.\ Rev.\ D {\bf 50} (1994) 1992.

\bibitem{dosch} H.G.~Dosch, C.~Ewerz and V.~Schatz,
Eur.\ Phys.\ J.\ C {\bf 24} (2002) 561


\bibitem{LT99} E.~Levin and K.~Tuchin,
{  Nucl. Phys.} {\bf B573} (2000) 833; { Nucl. Phys.} {\bf
A691} (2001) 779; { Nucl. Phys. } {\bf A693} (2001) 787.


\bibitem{AB01}
N.~Armesto and M.~Braun, { Eur. Phys. J.} {\bf C20} (2001) 517;
{\it ibid.} {\bf C22} (2001) 351.

\bibitem{Motyka} K.~Golec-Biernat, L.~Motyka, and A.M.~Sta\'sto,
{ Phys. Rev.} {\bf D65} (2002) 074037.

\bibitem{LL01}
E. Levin and M. Lublinsky, { Phys. Lett.} {\bf B521}  (2001) 233;
{ Eur. Phys. J.} {\bf C22} (2002) 647; M. Lublinsky, { Eur.
Phys. J.} {\bf C21} (2001) 513.

\bibitem{GBS03}
K. Golec-Biernat and A.M. Stasto, { Nucl.\ Phys.}\ {\bf B668}
(2003) 345.

\bibitem{Nestor03}
J.~L.~Albacete, N.~Armesto, A.~Kovner, C.~A.~Salgado and
U.~A.~Wiedemann,
{ Phys. Rev. Lett.} {\bf 92} (2004) 082001. 

\bibitem{Nestor04}
J.~L.~Albacete, N.~Armesto, J.~G.~Milhano, C.~A.~Salgado and
U.~A.~Wiedemann, {\it ``Numerical analysis of the
Balitsky-Kovchegov equation with running coupling: dependence of
the saturation scale on nuclear size and rapidity,"}
hep-ph/0408216.

\bibitem{SCALING}
E.~Iancu, K. Itakura, and L. McLerran, { Nucl. Phys.} {\bf A708}
(2002) 327.


\bibitem{MT02}
A. H. Mueller and D.N. Triantafyllopoulos, { Nucl. Phys.} {\bf
B640} (2002) 331.


\bibitem{DT02}D.N. Triantafyllopoulos, { Nucl. Phys.} {\bf B648} (2003) 293.

\bibitem{MP03}
S. Munier and R. Peschanski, { Phys. Rev. Lett.} {\bf 91} (2003)
232001 ; { Phys.\ Rev.}\  {\bf D69} (2004) 034008; {\it ibid.}
{\bf D70} (2004) 077503.


\bibitem{niko} N.N.~Nikolaev and B.G.~Zakharov,
Z.\ Phys.\ C {\bf 49} (1991) 607.

\end{thebibliography}
\end{document}